\documentclass[letterpaper,english,prd,preprint,nofootinbib,showpacs,preprintnumbers,fleqn,floatfix]{revtex4}
\usepackage[T1]{fontenc}
\usepackage[latin1]{inputenc}
\usepackage{graphicx}
\usepackage{amssymb}

\makeatletter

\usepackage{graphicx}
\usepackage{here}

\renewcommand{\vec}[1]{{\bf #1}}
\def\D0{D\O~}

\usepackage{babel}
\makeatother
\begin{document}

\newcommand{\prescr}[1]{{}^{#1}}

\newcommand{\ra}{\rightarrow}

\newcommand{\alphas}{\alpha_{s}}


\preprint{ANL-HEP-PR-04-97, CTEQ-0467, hep-ph/0410375}

\title{Transverse momentum resummation at small $x$ for the Tevatron and
LHC }

\author{Stefan Berge,$^{1}$%
\footnote{E-mail: berge@mail.physics.smu.edu%
} Pavel M. Nadolsky,$^{2}$%
\footnote{E-mail: nadolsky@hep.anl.gov%
} Fredrick I. Olness,$^{1}$%
\footnote{E-mail: olness@smu.edu%
} and C.-P.~Yuan$^{3}$%
\footnote{E-mail: yuan@pa.msu.edu%
}}

\affiliation{$^{1}$Department of Physics, Southern Methodist University, Dallas,
TX 75275-0175, U.S.A.\\
$^{2}$High Energy Physics Division, Argonne National Laboratory,\\
Argonne, IL 60439-4815, U.S.A. \\
$^{3}$Michigan State University, Department of Physics and Astronomy,
East Lansing, MI 48824-1116, U.S.A.}

\begin{abstract}
Analysis of semi-inclusive DIS hadroproduction suggests broadening
of transverse momentum distributions at $x$ below a few $10^{-3}$,
which can be modeled in the Collins-Soper-Sterman formalism by a modification
of impact-parameter-dependent parton densities. We discuss the consequences
of such a modification for the production of electroweak bosons at
hadron-hadron colliders. If substantial small-$x$ broadening is observed
in forward $Z^{0}$ boson production in the Tevatron Run-2, it will
strongly affect predicted $q_{T}$ distributions for $W^{\pm}$, $Z^{0}$,
and Higgs boson production at the Large Hadron Collider. 
\end{abstract}

\date{\today{}}

\maketitle

\section{Introduction\label{sec:Introduction}}

Moderately heavy particles with masses about 100 GeV can be copiously
produced in hadron collisions at TeV energies. Production and observation
of such particles is accompanied by intense hadronic activity, which
affects overall rates and differential cross sections. Radiation of
several hadrons of relatively low energy may produce a strong recoil
effect on the heavy particle. Calculations in perturbative quantum
chromodynamics (PQCD) may have to be reorganized to resum large logarithms
associated with multiple hadronic radiation.

Electroweak symmetry breaking in the standard model introduces vector
bosons $W^{\pm}$ and $Z^{0}$ with masses of about 80 and 91 GeV,
and a scalar Higgs boson $H^{0}$ with a likely mass below 250 GeV,
as suggested by global fits of electroweak data. Supersymmetric extensions
of the standard model require existence of several types of scalar
bosons, with the lightest naturally having a mass below about 140
GeV in the minimal supersymmetric standard model \cite{Degrassi:2002fi}.
Identification of heavy bosons relies on understanding of transverse
momentum distributions for heavy bosons and their decay products,
as well as for background processes. Short-lived bosons are commonly
detected by observing their decay into a pair of color-neutral particles
with large transverse momenta (of order of a half of the boson's mass)
in the laboratory reference frame.  Furthermore, the mass $M_{V}$
of the heavy boson (e.g., $W$ boson) can be determined from the transverse
momentum distribution of the decay products. 

When the boson's transverse momentum $q_{T}$ is much smaller than
the boson's virtuality $Q$, the calculation of the transverse momentum
distribution must evaluate an all-order sum of large logarithms $\ln^{n}(q_{T}/Q)$.
The resummation of small~$q_{T}$ logarithms can be realized using
one of several available methods; see \cite{Ellis:1998ii,Guffanti:2000ep,Kulesza:2001jc,Kulesza:2002rh,Ji:2004wu,Ji:2004xq}
for some recent references. As stated by Collins, Soper, and Sterman
(CSS) in Ref.~\cite{Collins:1985kg}, at moderate energies all large
logarithms are generated by a product of a soft (Sudakov) exponential
and special parton distribution functions in the impact parameter
($b$) space. The $b$-space resummed form factor can be evaluated
in PQCD at large $Q$ ($Q^{2}\gg\Lambda_{QCD}^{2}$) and small $b$
( $b^{2}\ll\Lambda_{QCD}^{-2}$). Its calculation is possible due
to the collinear factorization of hadronic cross sections, valid when
$Q$ is not small compared to the total energy $\sqrt{S}$ of the
hadronic collision. The CSS approach describes well the available
$q_{T}$ data from fixed-target Drell-Yan experiments and $Z$ boson
production at the Tevatron \cite{Landry:2002ix}. 

In this paper,%
\footnote{Our results were partly presented in Ref.~\cite{Berge:2004va} as
a contribution to the 3d Les Houches workshop {}``Physics at TeV
colliders'' \cite{Dobbs:2004bu}.%
} we discuss possible deviations from this established picture that
may occur in the new range of energies accessible to the Large Hadron
Collider (LHC). As we move from the 1.96 TeV $p\bar{p}$ collider
Tevatron to the 14 TeV $pp$ collider LHC, the typical fraction of
the collision energy going into production of moderately heavy bosons
decreases below one percent, leading to several new effects. Transverse
momentum distributions of $W$ and $Z$ bosons can be altered at small
momentum fractions $x$ by increased contributions from $qg$ and
$gg$ hard scattering, which tends to produce electroweak bosons with
larger transverse momenta as compared to the dominant process of $q\bar{q}$
scattering. One also expects enhancement of the logarithms $\ln(1/x)$
in the matrix elements of order $\alphas^{2}$ and beyond, the effect
not completely accounted for in the conventional $q_{T}$ resummation.
Such logarithms may be enhanced by a larger QCD coupling strength
$\alpha_{s}$ at $q_{T}$ less than a few GeV, even when they are
negligible in inclusive cross sections depending on one hard scale
$Q\gg1$ GeV. Finally, complex nonperturbative dynamics contributing
at $b\gtrsim1\mbox{ GeV}^{-1}$ may also depend on $x$. At present,
the magnitude of the $x$-dependent corrections to $W$, $Z$, and
Higgs boson production at the LHC energy is largely unknown, in part
because no experimental data on $q_{T}$ distributions is available
yet in Drell-Yan-like processes at $x$ of a few $10^{-3}$ or less. 

Resummation of $q_{T}$ logarithms at $q_{T}/Q\rightarrow0$ and $x\gg0$
is realized in the Dokshitzer-Gribov-Lipatov-Altarelli-Parisi (DGLAP)
picture of hadronic scattering \cite{Dokshitzer:1977sg,Gribov:1972ri,Gribov:1972rt,Altarelli:1977zs},
which assumes that the dominant parton emissions are strongly ordered
in transverse momenta. At asymptotically high collision energies ($x\sim Q/\sqrt{S}\rightarrow0$
with $q_{T}/Q$ fixed), the collinear DGLAP factorization is superseded
by the off-shell factorization in the Balitsky-Fadin-Kuraev-Lipatov
(BFKL) framework \cite{Balitsky:1978ic,Kuraev:1976ge,Kuraev:1977fs},
which sums up leading logarithms of $1/x$, while including $q_{T}$
logarithms at a finite order of $\alpha_{s}$ only. When neither DGLAP
nor BFKL dynamics dominates (for instance, when both $x$ and $q_{T}$
are small), methods for simultaneous summation of $q_{T}$ and $1/x$
logarithms (e.g., the Ciafaloni-Catani-Fiorani-Marchesini equation
\cite{Ciafaloni:1988ur,Catani:1990yc,Catani:1990sg,Marchesini:1995wr})
may be needed. Together with the increased magnitude of $qg$ and
$gg$ scattering, reduced convergence of the perturbation series in
the DGLAP framework may signal transition to the small-$x$ dynamical
regime at the existing colliders. 

Some of the transition phenomena may be already observed in part of
phase space at the $ep$ collider HERA. The analog of the Drell-Yan
process in lepton-nucleon deep inelastic scattering is semi-inclusive
production of hadronic final states, $e+p\stackrel{\gamma^{*}}{\longrightarrow}e+\mbox{{hadron(s)}}+X$.
The CSS resummation formalism can be applied to semi-inclusive deep
inelastic scattering (SIDIS) to describe the dependence on the Lorentz-invariant
transverse momentum $q_{T}$ \cite{Collins:1993kk,Meng:1996yn}. Refs.~\cite{Nadolsky:1999kb,Nadolsky:2000ky}
have compared predictions of $q_{T}$ resummation to the data for
the transverse energy flow in the current fragmentation region \cite{Aid:1995we,Adloff:1999ws}.
The experimentally observed $q_{T}$ distribution at $x$ below $10^{-2}$
becomes wider as $x$ decreases. This {}``broadening'' of $q_{T}$
distributions was modeled by including an extra $x$-dependent term
in the resummed form factor. To describe the data, the extra term
must grow quickly as $x\rightarrow0$. It provides a phenomenological
parametrization for all substantial factors responsible for $x$ dependence
beyond that included in the ${\cal O}(\alpha_{s})$ resummed $q_{T}$
cross section. For example, large corrections from hard scattering
of order ${\cal O}(\alpha_{s}^{2})$ substantially modify the SIDIS
cross section at large transverse momenta \cite{Kniehl:2004hf,Aurenche:2003by,Fontannaz:2004ev,Daleo:2004pn},
and such corrections are also likely important at small $q_{T}$ (not
discussed in the published studies). Apart from the hard-scattering
contributions, the dynamics of collinear radiation may change at moderately
small $x$ ($x<10^{-2}$) and small energy scales ($1/b<1-2$ GeV)
as a result of preasymptotic suppression of gluon splittings by BFKL
logarithms \cite{Ciafaloni:2003kd}. Both enhancement of perturbative
scattering at larger $q_{T}$ and preasymptotic BFKL suppression of
collinear radiation at smaller $q_{T}$ would result in broader $q_{T}$
distributions. 

While a systematic framework describing all $x$-dependent effects
is yet to be developed, some predictions can be made by exploiting
universality of soft and collinear radiative corrections, which dominate
the SIDIS energy flow data at $q_{T}<2-3$ GeV. For this reason, we
employ the phenomenological parametrization for the small-$q_{T}$
cross section found in SIDIS to predict the $x$ dependence of $q_{T}$
distributions at $x\lesssim10^{-2}$ at hadron-hadron colliders. 
The proposed new form of the Drell-Yan resummed form factor at $x\lesssim10^{-2}$
is given by\begin{equation}
\widetilde{W}(b,Q)=\widetilde{W}_{BLNY}(b,Q)e^{-\rho(x_{A})b^{2}-\rho(x_{B})b^{2}},\end{equation}
 where $\widetilde{W}_{BLNY}(b,Q)$ is the resummed form factor found
in the global fit \cite{Landry:2002ix} to the Drell-Yan data at larger
$x$, and the exponential $e^{-\rho(x_{A})b^{2}-\rho(x_{B})b^{2}}$
parametrizes the small-$x$ broadening.

New phenomena at small momentum fractions may have consequences for
precision measurements and searches for new physics. Modifications
in the transverse momentum distributions for $W$ and $Z$ bosons
will affect the measurements of the $W$ boson mass and width, as
well as the $W$ and $Z$ boson background in the search for new gauge
bosons, supersymmetry, particle compositeness, extra spatial dimensions,
etc. At the LHC, the $q_{T}$ broadening may affect detection of the
Higgs bosons by altering $q_{T}$ distributions of Higgs bosons and
the relevant QCD background. 

We examine possible small-$x$ effects in $W,$ $Z,$ and Higgs boson
production at the Tevatron and LHC. Sizable $q_{T}$ broadening may
occur at both colliders. The validity of our model can be tested in
the immediate future by the analysis of forward rapidity $q_{T}$
distributions in the Tevatron Run-2. An observation of signs of $q_{T}$
broadening at the Tevatron will suggest a large magnitude of this
effect at the LHC.

In Section~\ref{sec:Overview}, we propose a relationship between
the SIDIS and Drell-Yan $q_{T}$ distributions based on factorization
properties of the CSS resummed cross section. To describe the enhanced
$x$-dependent contributions, we include in the resummed form factor
an additional phenomenological term determined from the SIDIS data.
In Section~\ref{sec:Numerical-results}, we predict within this model
the $q_{T}$ distributions in production of $W$ and $Z$ bosons in
the forward region at the Tevatron and the central region at the LHC.
We also explore implications for light Higgs boson production via
gluon-gluon fusion at the LHC. The appendix reviews the rapidity coverage
at the existing hadron colliders.

\section{\label{sec:Overview}Relationship between the resummed cross sections
in Drell-Yan and semi-inclusive DIS processes}

\subsection{Resummed form factor in electroweak boson production\label{sub:Resum-large-x}}

We start by discussing production of the electroweak bosons at moderate
collision energies $\sqrt{S}$, when the ratio $Q/\sqrt{S}$ is large,
but not very close to unity. At such energies, the logarithms $\ln(x)$
or $\ln(1-x)$ of the partonic momentum fractions $x$ are of a moderate
magnitude in most of the events. This situation is typical for the
production of Drell-Yan lepton pairs in fixed-target experiments and
$W$ or $Z$ bosons at the Tevatron. 

At small transverse momenta $q_{T}$, the differential cross section
 takes the form \cite{Collins:1985kg}\begin{equation}
\frac{d\sigma}{dQ^{2}dydq_{T}^{2}}=\int_{0}^{\infty}\frac{bdb}{2\pi}\, J_{0}(q_{T}b)\,\widetilde{W}(b,Q,x_{A},x_{B})\,\,+\,\, Y(q_{T},Q,x_{A},x_{B}),\label{WYDY}\end{equation}
 where $y=\left(1/2\right)\ln\left[(E+p_{z})/(E-p_{z})\right]$ is
the rapidity of the vector boson, $x_{A,B}\equiv Qe^{\pm y}/\sqrt{S}$
are the Born-level partonic momentum fractions, and $J_{0}(q_{T}b)$
is the Bessel function. The integral is the Fourier-Bessel transform
of a form factor $\widetilde{W}(b,Q,x_{A},x_{B}$), given in impact
parameter ($b$) space. It contains an all-order sum of the logarithms
$\alpha_{s}^{n}\ln^{m}(q_{T}^{2}/Q^{2})$, which dominates in the
small-$q_{T}$ region. $Y(q_{T},Q,x_{A},x_{B})$ is the regular part,
defined as the difference of the fixed-order cross section and expansion
of the Fourier-Bessel integral to the same order of $\alpha_{s}$.
It is numerically small at $q_{T}\rightarrow0$.

$\widetilde{W}(b,Q,x_{A},x_{B})$ can be expressed as a product of
Born-level prefactors $\sigma_{ab}^{(0)}$, a Sudakov exponent $e^{-{\cal S}(b,Q)}$,
and $b$-space parton distributions ${\mathcal{\overline{P}}}_{a}^{(T)}(x,b)$:\begin{equation}
\widetilde{W}(b,Q,x_{A},x_{B})=\frac{\pi}{S}\sum_{a,b}\sigma_{ab}^{(0)}\, e^{-{\cal S}(b,Q)}\,\,{\mathcal{\overline{P}}}_{a}^{(T)}(x_{A},b)\,\,{\mathcal{\overline{P}}}_{b}^{(T)}(x_{B},b).\label{WCSS}\end{equation}
This form will be referred to as {}``the CSS representation''. It
has been stated as a factorization theorem in Ref.~\cite{Collins:1985kg}
and recently proved in Ref.~\cite{Collins:2004nx}.%
\footnote{The factorization for the resummed form factors in the Drell-Yan and
SIDIS processes has been also demonstrated recently within the framework
based on gauge-invariant $b$-dependent parton distributions \cite{Ji:2004wu,Ji:2004xq}.%
} The summation in Eq.~(\ref{WCSS}) is over the relevant parton flavors
$a$ and $b$. In the approximation of a vanishing width of the boson,
the non-zero Born-level prefactors are given by \begin{equation}
\sigma_{i\bar{j}}^{(0)}=\frac{\pi}{3}\sqrt{2}M_{W}^{2}G_{F}\left|V_{ij}\right|^{2}\delta(Q^{2}-M_{W}^{2})\end{equation}
in $W$ boson production for $i=u,d,...$, $\bar{j}=\bar{u},\bar{d},$...;
\begin{equation}
\sigma_{i\bar{i}}^{(0)}=\frac{\pi}{12}\sqrt{2}M_{Z}^{2}G_{F}\left(\left(1-4\left|e_{i}\right|\sin^{2}\theta_{w}\right)^{2}+1\right)\delta(Q^{2}-M_{Z}^{2})\end{equation}
in $Z$ boson production for $i=u,d,...$; and \begin{equation}
\sigma_{gg}^{(0)}=\frac{\alpha_{s}^{2}Q^{2}}{576\pi}\sqrt{2}G_{F}\delta(Q^{2}-M_{H}^{2})\label{sigma0Higgs}\end{equation}
in Higgs boson production. Here $M_{W},$ $M_{Z}$, and $M_{H}$ are
the masses of the $W,$ $Z,$ and Higgs bosons, $G_{F}$ is the Fermi
constant, $\theta_{w}=\sin^{-1}\left(1-M_{W}^{2}/M_{Z}^{2}\right)$
is the weak mixing angle, $V_{ij}$ is the Cabibbo-Kobayshi-Maskawa
matrix, and $e_{i}=2/3$ or $-1/3$ are the fractional quark charges
for up- or down-type quarks. The prefactor (\ref{sigma0Higgs}) for
Higgs boson production is derived by using the effective $ggH$ vertex
in the approximation of infinitely heavy top quarks \cite{Ellis:1975ap,Shifman:1979eb,Dawson:1990zj,Djouadi:1991tk}.
This approximation works well for the Higgs bosons that are lighter
than the top quark.

The Sudakov term ${\cal S}(b,Q)$ is an integral of the functions
${\cal A}\left(\alpha_{s}(\bar{\mu})\right)$ and ${\cal B}\left(\alpha_{s}(\bar{\mu})\right)$,
which can be calculated in PQCD when $b$ is small ($b^{2}\ll\Lambda_{QCD}^{-2}$):\begin{equation}
{\cal S}(b,Q)=\int_{b_{0}^{2}/b^{2}}^{Q^{2}}\frac{d\bar{\mu}^{2}}{\bar{\mu}^{2}}\left[{\cal A}\left(\alpha_{s}(\bar{\mu})\right)\ln\frac{Q^{2}}{\bar{\mu}^{2}}+{\cal B}\left(\alpha_{s}(\bar{\mu})\right)\right].\label{Spert}\end{equation}
$b_{0}\equiv2e^{-\gamma_{E}}$ is a constant parameter involving the
Euler constant, $\gamma_{E}=0.577...$ . Approximation of ${\cal S}(b,Q)$
at a finite order of $\alpha_{s}$ yields {}``the perturbative Sudakov
factor'' ${\cal S}_{P}(b,Q)$. The perturbative coefficients ${\cal A}^{(i)}$
and ${\cal B}^{(i)}$ for the functions ${\cal A}\left(\alpha_{s}(\bar{\mu})\right)$
and ${\cal B}\left(\alpha_{s}(\bar{\mu})\right)$ up to the next-to-next-to-leading
order are published in Refs.~\cite{Collins:1985kg,Kodaira:1982nh,Davies:1984hs}
for the Drell-Yan process, and Refs.~\cite{Catani:1988vd,Kauffman:1991jt,Yuan:1992we,deFlorian:2000pr,Catani:2000vq}
for Higgs boson production.%
\footnote{Equations in this paper correspond to the {}``canonical'' choice
\cite{Collins:1985kg} of the momentum scales.%
} 

The form factor (\ref{WCSS}) depends on the light-cone momentum fractions
$x_{A}$ and $x_{B}$ through the functions ${\mathcal{\overline{P}}}_{a}^{(T)}(x,b)$.
In the CSS representation, we distinguish between the functions ${\mathcal{\overline{P}}}_{a}^{(T)}(x,b)$
and ${\mathcal{\overline{P}}}_{a}^{(S)}(x,b)$ appearing in reactions
with timelike $(T)$ and spacelike $(S)$ electroweak bosons. The
relationship between ${\mathcal{\overline{P}}}_{a}^{(T)}(x,b)$ (relevant
to the Drell-Yan process) and ${\mathcal{\overline{P}}}_{a}^{(S)}(x,b)$
(relevant to SIDIS) is discussed later in this section. Expansion
of ${\mathcal{\overline{P}}}_{a}^{(T)}(x,b)$ in leading powers of
$b$ at $b\rightarrow0$ gives \cite{Collins:1982uw} \begin{equation}
{\mathcal{\overline{P}}}_{a}^{(T)}(x,b)=\sum_{c}\int_{x}^{1}\frac{d\xi}{\xi}{\cal C}_{a/c}^{in\,(T)}\left(\xi,b,\alpha_{s}\left(\frac{b_{0}}{b}\right)\right)f_{c}\left(\frac{x}{\xi},\frac{b_{0}}{b}\right)+...\equiv\left[{\cal C}_{a/c}^{in\,(T)}\otimes f_{c}\right](x,b)+...\,.\label{PDY}\end{equation}
 Here ${\cal C}_{a/c}^{in\,(T)}\left(x,b,\alpha_{s}(b_{0}/b)\right)$
are the Wilson coefficient functions for incoming ({}``in'') partons,
and $f_{c}(x,\, b_{0}/b)$ are the conventional parton distributions
(integrated over the parton's transverse momentum $\vec{k}_{T}$).
The ellipses denote the terms that are suppressed by positive powers
of $b$. The renormalization and factorization scales on the right-hand
side of Eq.~(\ref{PDY}) are equal to $b_{0}/b$. Whenever the series
in powers of $\alpha_{s}(b_{0}/b)$ and $b$ in Eq.~(\ref{PDY})
converge, ${\mathcal{\overline{P}}}_{a}^{(T)}(x,b)$ can be approximated
by neglecting the positive powers of $b$, evaluating ${\cal C}_{a/c}^{in\,(T)}$
at a finite order of $\alpha_{s}$, and using a parametrization for
$f_{c}(x,\mu)$ from a phenomenological global fit. Estimates for
${\cal C}_{a/c}^{in\,(T)}$ of order ${\cal O}(\alpha_{s})$ can be
found in Refs.~\cite{Collins:1985kg,Kauffman:1991jt,Yuan:1992we}. 

Contributions from nonperturbative impact parameters ($b\gtrsim1\mbox{\, GeV}{}^{-1}$)
are suppressed at $Q\rightarrow\infty$ by the shape of $\widetilde{W}(b,Q,x_{A},x_{B}$)
and oscillations of $J_{0}(q_{T}b)$. Nonetheless, mild sensitivity
to the large-$b$ behavior remains in $W$ and $Z$ boson production
when $q_{T}$ is of order a few GeV. The factorization of $\widetilde{W}(b,Q,x_{A},x_{B})$
predicts universality of the nonperturbative terms within classes
of similar processes, such as production of Drell-Yan pairs, $W$,
and $Z$ bosons. While the nonperturbative terms cannot be calculated
yet from the first principles, they can be introduced in the calculation
by using one of the available models. The present Drell-Yan and $Z$
boson data at $x$ above a few $10^{-2}$ is consistent with universality,
as it was demonstrated recently \cite{Landry:2002ix} within the {}``$b_{*}$''
model \cite{Collins:1982va,Collins:1985kg} for the nonperturbative
terms. The {}``$b_{*}$'' model parametrizes the resummed form factor
as\begin{equation}
\widetilde{W}(b,Q,x_{A},x_{B})=\widetilde{W}^{pert}(b_{*},Q,x_{A},x_{B})\, e^{-{\mathcal{S}}_{NP}(b,Q;b_{*})},\label{Wbstar}\end{equation}
where\begin{equation}
\widetilde{W}^{pert}(b_{*},Q,x_{A},x_{B})\equiv\frac{\pi}{S}\sum_{a,b}\sigma_{ab}^{(0)}\left[{\mathcal{C}}_{a/c}^{in\,(T)}\otimes f_{c}\right]\left(x_{A},\frac{b_{0}}{b_{*}}\right)\,\left[{\mathcal{C}}_{b/d}^{in\,(T)}\otimes f_{d})\right]\left(x_{B},\frac{b_{0}}{b_{*}}\right)\, e^{-{\mathcal{S}}_{P}(b_{*},Q)}\label{Wpert}\end{equation}
 is the perturbative approximation for $\widetilde{W}(b,Q,x_{A},x_{B}),$
evaluated as a function of $b_{*}\equiv b(1+b^{2}/b_{max}^{2})^{-1/2}$
for $b_{max}\sim1\mbox{\, GeV}{}^{-1}$. The nonperturbative Sudakov
function,\begin{equation}
{\cal S}_{NP}(b,Q;b_{*})\equiv-\ln\left[\frac{\widetilde{W}(b,Q,x_{A},x_{B})}{\widetilde{W}^{pert}(b_{*},Q,x_{A},x_{B})}\right],\label{SNP}\end{equation}
is determined from the experimental data. If the flavor dependence
of large-$b$ contributions is neglected (a reasonable first approximation
in reactions dominated by the scattering of $u$ and $d$ quarks),
${\mathcal{S}}_{NP}(b,Q;b_{*})$ is straightforwardly related to the
true Sudakov factor ${\cal S}(b,Q)$ and $b$-dependent parton distributions
${\mathcal{\overline{P}}}_{a}^{(T)}(x_{A},b)$ in Eq.~(\ref{WCSS}):\begin{equation}
{\cal S}_{NP}(b,Q;b_{*})\approx{\cal S}(b,Q)-{\cal S}_{P}(b_{*},Q)-\ln\left(\frac{{\mathcal{\overline{P}}}_{a}^{(T)}(x_{A},b)}{{\mathcal{\overline{P}}}_{a}^{(T)}(x_{A},b_{*})}\right)-\ln\left(\frac{{\mathcal{\overline{P}}}_{b}^{(T)}(x_{B},b)}{{\mathcal{\overline{P}}}_{b}^{(T)}(x_{B},b_{*})}\right).\label{SNPST}\end{equation}
This representation realizes the generic form for ${\cal S}_{NP}(b,Q;b_{*})$
found in Ref.~\cite{Collins:1985kg},\[
{\cal S}_{NP}(b,Q;b_{*})=g_{S}(b,\ln Q;b_{*})+g_{a}(b,x_{A};b_{*})+g_{b}(b,x_{B};b_{*}),\]
where $g_{S},$ $g_{a}$, and $g_{b}$ are some functions of the shown
arguments. Several analyses have been carried out in the past in order
to constrain ${\cal S}_{NP}(b,Q;b_{*})$ from the Drell-Yan data \cite{Davies:1984sp,Ladinsky:1993zn,Ellis:1997sc,Landry:1999an,Landry:2001qr}.
In the present work, we use the BLNY parametrization ${\mathcal{S}}_{NP}^{BLNY}$
\cite{Landry:2002ix}, given by \begin{equation}
{\mathcal{S}}_{NP}^{BLNY}(b,Q;b_{*})=\left[0.21+0.68\ln\left(\frac{Q}{3.2\mbox{\, GeV}}\right)-0.126\ln{(100x_{A}x_{B})}\right]b^{2}\label{BLNY}\end{equation}
\noindent for $b_{max}=0.5\mbox{\, GeV}{}^{-1}$. This choice of
parameters yields good agreement with the Drell-Yan data and is convenient
for comparison with the $q_{T}$ fit in SIDIS, which was also performed
for $b_{max}=0.5\mbox{ GeV}^{-1}$. Alternative forms for the parametrization
of $\widetilde{W}(b,Q,x_{A},x_{B})$ at large $b$ have been also
considered \cite{Qiu:2000hf,Kulesza:2002rh,Guffanti:2000ep}. Studies
\cite{Korchemsky:1999kt,Tafat:2001in} of the general structure of
the large-$b$ contributions with the methods of infrared renormalon
analysis have demonstrated, in particular, that the leading power-suppressed
contribution is quadratic in $b$ (Gaussian), \emph{i.e.}, that ${\cal S}_{NP}\sim b^{2}$,
as in Eq.~(\ref{BLNY}).

Since $x_{A}x_{B}=Q^{2}/S,$ ${\mathcal{S}}_{NP}^{BLNY}(b,Q;b_{*})$
is independent of the boson's rapidity. It will be used in Section~\ref{sec:Numerical-results}
to predict the $W$ and $Z$ boson resummed cross sections in the
case when all rapidity dependence is contributed by $\left[{\mathcal{C}}_{a/c}^{in\,(T)}\otimes f_{c}\right]\left(x,b_{0}/b_{*}\right)$.
By comparing Eqs.~(\ref{SNPST}) and (\ref{BLNY}), we find that
the logarithmic combination of ${\mathcal{\overline{P}}}_{a}^{(T)}(x,b)$,\begin{equation}
\ln\left(\frac{{\mathcal{\overline{P}}}_{a}^{(T)}(x,b)}{{\mathcal{\overline{P}}}_{a}^{(T)}(x,b_{*})}\right)\approx\ln\left(\frac{{\mathcal{\overline{P}}}_{a}^{(T)}(x,b)}{\left[{\mathcal{C}}_{a/c}^{in\,(T)}\otimes f_{c}\right]\left(x,\frac{b_{0}}{b_{*}}\right)}\right)\equiv c^{(T)}(x,b;b_{*}),\label{ST}\end{equation}
is a slowly varying function of $x$ (denoted by $c^{(T)}(x,b;b_{*})$).
Therefore, \begin{equation}
{\mathcal{\overline{P}}}_{a}^{(T)}(x,b)\approx\left[{\mathcal{C}}_{a/c}^{in\,(T)}\otimes f_{c}\right]\left(x,\frac{b_{0}}{b_{*}}\right)e^{c^{(T)}(x,b;b_{*})},\label{PTlargex}\end{equation}
when $x$ is larger than a few percent. 

The nonperturbative function ${\mathcal{S}}_{NP}(b,Q;b_{*})$ may
take a different form in gluon-dominated Higgs boson production. Fortunately,
the sensitivity to ${\mathcal{S}}_{NP}(b,Q;b_{*})$ is suppressed
in gluon-initiated channels at the LHC by a larger color factor associated
with the perturbative leading-logarithm contributions. In Section~\ref{sub:LHC-Higgs},
we quantitatively estimate the effect of the uncertainty in ${\mathcal{S}}_{NP}(b,Q;b_{*})$
on the Higgs boson cross section.

\subsection{Resummed form factor at small $x$\label{sub:Resum-small-x}}

As $x\rightarrow0$, the approximation (\ref{PTlargex}) for ${\mathcal{\overline{P}}}_{a}^{(T)}(x,b)$
may be altered because of the reduced convergence of the perturbation
series at intermediate and large impact parameters. The factorized
expression (\ref{WCSS}) for $\widetilde{W}(b,Q,x_{A},x_{B})$ is
derived from the requirements of renormalization- and gauge-group
invariance, \emph{}and does not rely on the perturbative convergence
for the individual components, including ${\mathcal{\overline{P}}}_{a}^{(T)}(x,b)$.
Similar factorization holds for the resummed form factor in semi-inclusive
deep inelastic scattering (SIDIS), $e+A\stackrel{\gamma^{*}}{\longrightarrow}e+B+X$
\cite{Collins:1993kk,Meng:1996yn,Nadolsky:1999kb,Ji:2004wu}, where
$A$ and $B$ are the initial- and final-state hadrons, respectively.
SIDIS probes the $x$-dependence of the \emph{spacelike} distributions
${\mathcal{\overline{P}}}_{j}^{(S)}(x,b)$ for \emph{quark} flavors
($j=u,d,...)$, which differ from the timelike quark distributions
${\mathcal{\overline{P}}}_{j}^{(T)}(x,b)$ by a factor $r(b)$: \begin{equation}
\overline{{\cal P}}_{j}^{(T)}(x,b)=r(b)\overline{{\cal P}}_{j}^{(S)}(x,b).\label{PHHP}\end{equation}
This relationship is derived in Section~\ref{sub:From-SIDIS-to-DY},
which also shows that $r(b)$ is independent of $x$. 

The experimentally observed hadronic energy flow in SIDIS \cite{Aid:1995we,Adloff:1999ws}
is compatible with the following form of ${\mathcal{\overline{P}}}_{j}^{(S)}(x,b)$
\cite{Nadolsky:1999kb,Nadolsky:2000ky} at $10^{-4}\lesssim x\lesssim1$:\begin{equation}
10^{-4}\lesssim x\lesssim1:\,{\mathcal{\overline{P}}}_{j}^{(S)}(x,b)\approx\left[{\mathcal{C}}_{j/a}^{in\,(S)}\otimes f_{a}\right](x,b_{*})\,\, e^{-0.013b^{2}/x+c^{(S)}(x,b;b_{*})},\label{PSsmallx}\end{equation}
with\begin{equation}
{\mathcal{C}}_{j/a}^{in\,(S)}(x,b_{*})=\frac{1}{r(b_{*})}{\mathcal{C}}_{j/a}^{in\,(T)}(x,b_{*}).\label{CinSCinT}\end{equation}
 The exponential $\exp\left(-0.013b^{2}/x\right)$ results in the
additional $x$-dependent broadening of the $q_{T}$ distributions.
The power of the exponential, $0.013b^{2}/x$, is small at $x\gg10^{-2}$
and rapidly grows at $10^{-4}<x<10^{-2}$. The SIDIS fit does not
constrain ${\mathcal{\overline{P}}}_{j}^{(S)}(x,b)$ at $x<10^{-4}$.
The $1/x$ growth of the exponential power may be modified at very
small $x$ by the turn-on of other scattering mechanisms, notably
BFKL resummation and saturation. The remaining terms (collectively
denoted as $c^{(S)}(x,b;b_{*})$) do not depend strongly on $x$ and
can be numerically important only at large $x$ and $b$. The behavior
of ${\mathcal{\overline{P}}}_{j}^{(S)}(x,b)$ in the transition region
$x\sim10^{-2}$ and precise form of $c^{(S)}(x,b;b_{*})$ are not
well determined by the SIDIS fit because of the limited accuracy of
the data. 

According to Eq.~(\ref{PHHP}), the broadening exponential must also
be present in the timelike distributions ${\mathcal{\overline{P}}}_{j}^{(T)}(x,b)$
and dominate at moderately small $x$: \begin{equation}
10^{-4}<x<10^{-2}:\,{\mathcal{\overline{P}}}_{j}^{(T)}(x,b)\approx\left[{\mathcal{C}}_{j/a}^{in\,(T)}\otimes f_{a}\right](x,b_{*})\,\, e^{-0.013b^{2}/x+c^{(T)}(x,b;b_{*})}.\label{PTsmallx}\end{equation}
The functions $c^{(T)}(x,b;b_{*})$ and $c^{(S)}(x,b;b_{*})$ are
in principle related by \begin{equation}
c^{(T)}(x,b;b_{*})=c^{(S)}(x,b;b_{*})+\ln\left[\frac{r(b)}{r(b_{*})}\right].\label{cTcS}\end{equation}
We find that the large-$x$ form (\ref{Wbstar}) of the resummed form
factor is quantitatively known from the fits to the Drell-Yan and
Tevatron Run-1 $Z$ boson data, while the leading behavior at $x<10^{-2}$
can be inferred from SIDIS. What is not precisely known is the behavior
of ${\mathcal{\overline{P}}}_{j}^{(T)}(x,b)$ in the transition region
around $x\sim10^{-2}$. To estimate the possible impact on the Tevatron
and LHC observables, we interpolate ${\mathcal{\overline{P}}}_{j}^{(T)}(x,b)$
between the large-$x$ parametrization (\ref{PTlargex}) and small-$x$
parametrization (\ref{PTsmallx}) by using a smooth trial function
$\rho(x)$:\begin{equation}
10^{-4}\lesssim x<1:\,{\mathcal{\overline{P}}}_{j}^{(T)}(x,b)\approx\left[{\mathcal{C}}_{j/a}^{in\,(T)}\otimes f_{a}\right](x,b_{*})\,\, e^{-\rho(x)b^{2}+c^{(T)}(x,b;b_{*})}.\label{PTallx}\end{equation}
 We choose $\rho(x)$ as\begin{equation}
\rho(x)=c_{0}\left(\sqrt{\frac{1}{x^{2}}+\frac{1}{x_{0}^{2}}}-\frac{1}{x_{0}}\right),\label{eq:ax}\end{equation}
where $c_{0}$ controls the magnitude of the broadening for a given
$x$, and $x_{0}$ is a characteristic value of $x$ below which $\rho(x)$
becomes non-negligible. In the limits $x\gg x_{0}$ and $x\ll x_{0}$,
Eq.~(\ref{PTallx}) reduces to the large-$x$ form (\ref{PTlargex})
and small-$x$ form (\ref{PTsmallx}), respectively. As a result of
the new form for ${\mathcal{\overline{P}}}_{j}^{(T)}(x,b)$, the BLNY
representation (\ref{Wbstar}) is multiplied by $e^{-\rho(x_{A})b^{2}-\rho(x_{B})b^{2}}$
: \begin{eqnarray}
\widetilde{W}(b,Q,x_{A},x_{B}) & = & \frac{\pi}{S}\sum_{a,b}\sigma_{ab}^{(0)}\left[{\mathcal{C}}_{a/c}^{in\,(T)}\otimes f_{c}\right]\left(x_{A},\frac{b_{0}}{b_{*}}\right)\,\left[{\mathcal{C}}_{b/d}^{in\,(T)}\otimes f_{d})\right]\left(x_{B},\frac{b_{0}}{b_{*}}\right)\nonumber \\
 & \times & e^{-{\mathcal{S}}_{P}(b_{*},Q)-S_{NP}^{BLNY}(b,Q;b_{*})}\, e^{-b^{2}\rho(x_{A})-b^{2}\rho(x_{B})}.\label{eq:full}\end{eqnarray}
 Eq.~(\ref{Wbstar}) is restored from Eq.~(\ref{eq:full}) when
both $x_{A}$ and $x_{B}$ are much larger than $x_{0}.$ The Gaussian
exponential $e^{-\rho(x)b^{2}}$approximately preserves the integral
of the $q_{T}$ distribution, and the observables inclusive in $q_{T}$
remain essentially unaffected. To reproduce the leading small-$x$
behavior in Eq.~(\ref{PTsmallx}), our quantitative estimates for
$W$ and $Z$ boson cross sections in Section~\ref{sec:Numerical-results}
are made for $c_{0}=0.013$ and $x_{0}=0.005$. We will also consider
variations around these values to examine sensitivity of the Tevatron
$W$ and $Z$ production to various broadening scenarios. The cross
sections for Higgs boson production are computed for two values of
$c_{0}$, $0.013$ and $0.026$, in order to evaluate the uncertainty
due to the unknown small-$x$ form of the gluon distribution ${\mathcal{\overline{P}}}_{g}^{(T)}(x,b)$
(not constrained by the SIDIS data).

\subsection{Resummed $z$ flow\label{sub:z-flow}}

In the remainder of this section, we review the resummation in SIDIS
$e(k)+A(p_{A})\stackrel{\gamma^{*}(q)}{\longrightarrow}e(k')+B(p_{B})+X$
and its relationship to the resummation in the Drell-Yan process.
The $x$ dependence of the distribution ${\mathcal{\overline{P}}}_{a}^{(S)}(x,b)$
can be determined from the data on the hadronic $z$-flow $\Sigma_{z}$,
defined by \cite{Peccei:1979tc,Peccei:1980kn,Dechantsreiter:1981ge}\begin{equation}
\Sigma_{z}\equiv\sum_{B}\int_{z_{min}}^{1}z\ \ \sigma(e+A\rightarrow e+B+X)\ \  dz.\end{equation}
The $z$-flow is obtained from the product of the cross section and
final-state light-cone variable $z=(p_{B}\cdot p_{A})/(q\cdot p_{A})$
by integrating over $z$ and summing over all final-state hadrons
$B$. It is related to the pseudorapidity ($\eta$) distribution of
the transverse energy flow $\langle E_{T}\rangle$ in the center-of-mass
frame (c.m.) of the virtual photon and initial-state hadron $A$ (with
$\gamma^{*}$ moving in the $+z$ direction): \begin{equation}
q_{T}^{2}\frac{d\Sigma_{z}}{dxdQ^{2}dq_{T}}\equiv\frac{d\langle E_{T}\rangle}{dxdQ^{2}d\eta},\end{equation}
where $Q^{2}=-q^{2}>0$, and $x\equiv Q^{2}/(2p_{A}\cdot q)$. The
Lorentz-invariant transverse momentum $q_{T}$ is a function of $\eta$,
$q_{T}=Q(x^{-1}-1)^{1/2}e^{-\eta}$ \cite{Nadolsky:1999kb}. The resummation
applies in the limit $q_{T}\rightarrow0$, which corresponds to the
current fragmentation region in the photon-hadron c.m. frame ($\eta\rightarrow+\infty$).
In this limit, the enhanced soft and collinear logarithms in the theory
prediction must be summed to all orders.

Similarly to the Drell-Yan case of Eq.~(\ref{WYDY}), the resummed
$z$-flow is given by a combination of the Fourier-Bessel transform
integral and finite term $Y_{z}$:\begin{equation}
\frac{d\Sigma_{z}}{dxdQ^{2}dq_{T}^{2}}=\int_{0}^{\infty}\frac{bdb}{2\pi}\,\, J_{0}(q_{T}b)\,\,\widetilde{W}_{z}(b,Q,x)+Y_{z}(q_{T},Q,x).\label{WYSIDIS}\end{equation}
Under HERA conditions, the finite piece $Y_{z}(q_{T},Q,x)$ is small
at $q_{T}<2-3$ GeV. The analysis of the SIDIS $z$-flow at such $q_{T}$
probes universal soft and collinear radiation described by the resummed
Fourier-Bessel integral. The form factor $\widetilde{W}_{z}(b,Q,x)$
is structurally close to the resummed form factor (\ref{WCSS}) in
the Drell-Yan process:

\begin{equation}
\widetilde{W}_{z}(b,Q,x)=\frac{\pi}{S_{eA}}\sum_{j=u,\bar{u},d,\bar{d},...}\sigma_{j}^{(0)}e^{-{\cal S}(b,Q)}{\cal C}_{z}^{out}(b)\overline{{\cal P}}_{j}^{(S)}(x,b),\label{WzCSS}\end{equation}
where $S_{eA}\equiv(k+p_{A})^{2}$,\begin{equation}
\sigma_{j}^{(0)}\equiv\frac{8\pi\alpha_{EM}^{2}}{S_{eA}x^{2}}\left(1-\frac{2}{y_{DIS}}+\frac{2}{y_{DIS}^{2}}\right)e_{j}^{2},\end{equation}
 and $y_{DIS}\equiv Q^{2}/(xS_{eA}).$ The structure of the Sudakov
exponent ${\cal S}(b,Q)$ is displayed in Eq.~(\ref{Spert}). The
spacelike distributions $\overline{{\cal P}}_{j}^{(S)}(x,b)$ factorize
at small $b$ as \begin{equation}
\left.{\mathcal{\overline{P}}}_{j}^{(S)}(x,b)\right|_{b^{2}\ll\Lambda_{QCD}^{-2}}=\left[{\cal C}_{j/a}^{in\,(S)}\otimes f_{a}\right](x,b),\label{PSIDIS}\end{equation}
where ${\cal C}_{j/a}^{in\,(S)}\left(x,b,\alpha_{s}(b_{0}/b)\right)$
are the spacelike Wilson coefficient functions for incoming partons.
${\cal C}_{z}^{out}(b)$ is a residual from the summation over the
hadronic final states:\begin{equation}
{\cal C}_{z}^{out}(b)=1+\frac{\alpha_{s}(b_{0}/b)}{\pi}C_{F}\left(-1-\frac{\pi^{2}}{3}\right)+...\,.\label{Coutzb}\end{equation}

The right-hand side of Eq.~(\ref{WzCSS}) depends on $x$ via $\overline{P}_{j}^{(S)}(x,b)$.
Nonperturbative contributions to the $z$-flow can be parametrized
using the $b_{*}$ prescription: \begin{equation}
\widetilde{W}_{z}(b,Q,x)=\frac{\pi}{S_{eA}}\sum_{j=u,\bar{u},d,\bar{d},...}\sigma_{j}^{(0)}{\cal C}_{z}^{out}(b_{*})\left[{\cal C}_{j/a}^{in}\otimes f_{a}\right](x,b_{*})e^{-{\cal S}_{P}(b_{*},Q)-{\cal S}_{NP}^{z}(b,Q,x;b_{*})},\label{Wzbstar}\end{equation}
where ${\cal S}_{NP}^{z}(b,Q,x;b_{*})$ was found from an ${\cal O}(\alpha_{s})$
fit \cite{Nadolsky:2000ky} to the data \cite{Aid:1995we,Adloff:1999ws}
at $8\nolinebreak\times\nolinebreak10^{-5}\leq\nolinebreak\langle x\rangle\le\nolinebreak0.11$
and $10\nolinebreak\le\nolinebreak\langle Q^{2}\rangle\nolinebreak\le\nolinebreak2200\mbox{{\, GeV}}{}^{2}$.
The leading small-$x$ behavior of ${\cal S}_{NP}^{z}(b,Q,x;b_{*})$
is consistent with inverse power dependence: \begin{eqnarray}
{\cal S}_{NP}^{z}(b,Q,x;b_{*}) & \approx & \frac{c_{0}b^{2}}{x^{p}}+...\,,\label{SNPzpar2}\end{eqnarray}
 The best fit in Ref.~\cite{Nadolsky:2000ky} quotes $c_{0}=0.013$
and $p=1$. The $x$-dependence shown in Eq.~(\ref{SNPzpar2}) describes,
in particular, the data at $q_{T}<2-3$ GeV, where the uncertainties
due to the matching (switching) between the $W$ and $Y$ terms are
small. By comparing two forms (\ref{WzCSS}) and (\ref{Wzbstar})
for $\widetilde{W}_{z}(b,Q,x)$, we find that the leading small-$x$
term is a part of $\overline{P}_{j}^{(S)}(x,b)$, in agreement with
Eq.~(\ref{PSsmallx}): \begin{equation}
{\mathcal{\overline{P}}}_{j}^{(S)}(x,b)\approx\left[{\mathcal{C}}_{j/a}^{in\,(S)}\otimes f_{a}\right](x,b_{*})\, e^{-\frac{0.013b^{2}}{x}+c^{(S)}(x,b;b_{*})}.\label{PSjxb}\end{equation}

Theoretical uncertainties in the parametrization of $\overline{{\cal P}}_{j}^{(S)}(x,b)$
due to final-state fragmentation, matching, and higher-order corrections
to the $Y$-term are reduced by our choice of the experimental observable
($z$-flow) and kinematical region ($q_{T}$ less than a few GeV).
The exponential factor $e^{-0.013b^{2}/x}$ parametrizes the full
contribution of higher-order radiative corrections of diverse wavelengths
to ${\mathcal{\overline{P}}}_{j}^{(S)}(x,b)$. It indicates that the
SIDIS $q_{T}$ distributions tend to become wider at smaller $x$,
and it attributes the observed broadening to scattering at $b^{2}\gtrsim0.013/x$.
At $x=10^{-2}\,(10^{-4})$, this inequality translates into $b_{0}/b<b_{0}(x/0.013)^{-1/2}\approx1.3\,(13)$
GeV, i.e., only scattering at low and intermediate energy scales is
affected. Several intense mechanisms beyond ${\cal O}(\alpha_{s})$
may turn on at such small $x$ and $1/b$ and cause the broadening.
The rate may be increased at larger $q_{T}$ ($1/b>1-2$ GeV) by enhanced
contributions from $qg$ and $gg$ hard scattering, as it happens
in ${\cal O}(\alpha_{s}^{2})$ SIDIS cross sections at $p_{T}=zq_{T}\sim Q$
\cite{Kniehl:2004hf,Aurenche:2003by,Fontannaz:2004ev,Daleo:2004pn}.
The rate may be reduced at smaller $q_{T}$ ($1/b<1-2$ GeV) in the
probed range of $x$ because of the suppression of collinear splittings
of gluons by the logarithms $\ln(1/x)$ at higher orders of $\alpha_{s}$
\cite{Ciafaloni:2003kd}. In addition to these leading-power (logarithmic
in $b$) contributions, the $x$ dependence may also arise from the
unknown nonperturbative terms in ${\mathcal{\overline{P}}}_{j}^{(S)}(x,b)$
proportional to the positive powers of $b$. In the absence of a systematic
computation of all such effects in the resummation formalism, the
phenomenological broadening exponential serves as an approximate model
of their total impact in the limited range $10^{-4}<x<10^{-2}$ covered
by the H1 data. The form of ${\mathcal{\overline{P}}}_{j}^{(S)}(x,b)$
is unknown at $x$ outside of the H1 data range, so that the parametrization
(\ref{PSjxb}) may not be extrapolated to $x$ below $10^{-4}$.

\subsection{Relationship between resummation in DY and SIDIS\label{sub:From-SIDIS-to-DY}}

The remaining step in the derivation of Eq.~(\ref{eq:full}) is the
relationship (\ref{PHHP}) between the timelike distributions $\overline{{\cal P}}_{j}^{(T)}(x,b)$
and spacelike distributions $\overline{{\cal P}}_{j}^{(S)}(x,b)$.
This relationship follows from an elementary representation for spacelike
($I=S$) and timelike ($I=T$) distributions ${\mathcal{\overline{P}}}_{j}^{(I)}(x,b)$
\cite{Collins:1982va}:\begin{equation}
\overline{{\cal P}}_{j}^{(I)}(x,b)=\left|{\cal H}_{j}^{(I)}(b_{0}/b)\right|\widetilde{U}(b_{0}/b)^{1/2}\widehat{{\cal P}}_{j}(x,b).\label{PHUP}\end{equation}
The function ${\cal H}_{j}^{(I)}(b_{0}/b)=1+\left(\alpha_{s}(b_{0}/b)/\pi\right){\cal H}_{j}^{(1,\, I)}+...$
is composed of highly virtual corrections to the hard scattering vertex.
The function $\widetilde{U}(b_{0}/b)^{1/2}$ collects the soft subgraphs
that are attached by gluon propagators to ${\cal H}_{j}^{(I)}(b_{0}/b)$.
It is the same for $I=S$ and $I=T$. $\widehat{{\cal P}}_{j}(x,b)$
is related to the $(n-2)$-dimensional Fourier-Bessel transform of
the unintegrated ($\vec{k}_{T}$-dependent) parton distribution function
${\cal P}_{j}(x,\vec{k}_{T},\zeta)$, \begin{eqnarray}
 &  & \widehat{{\cal P}}_{j}(x,b)=\lim_{\zeta\rightarrow\infty}\Biggl\{ e^{{\cal S}'(b,\zeta)}\int d^{n-2}\vec{k}_{T}e^{i\vec{k}_{T}\cdot\vec{b}}{\cal {P}}_{j}^{in}(x,\vec{k}_{T},\zeta)\Biggr\}.\label{limzeta}\end{eqnarray}
${\cal P}_{j}(x,\vec{k}_{T},\zeta)$ is given in terms of the quark
field $\psi_{j}$ by\begin{equation}
{\cal P}_{j}(x,\vec{k}_{T},\zeta)=\overline{\sum_{spin}}\,\overline{\sum_{color}}\int\frac{dy^{-}d^{2}\vec{y}_{T}}{(2\pi)^{3}}e^{-ixp^{+}y^{-}+i\vec{k}_{T}\cdot\vec{y}_{T}}\langle p|\bar{\psi}_{j}(0^{+},y^{-},\vec{y}_{T})\frac{\gamma^{+}}{2}\psi_{j}(0)|p\rangle,\end{equation}
 in a gauge $n\cdot{\cal A}=0$ with $n^{2}<0$. It depends on the
gauge through the parameter $\zeta\nolinebreak\equiv\nolinebreak(p\cdot n)/|n^{2}|$,
where $p^{\mu}$ is the momentum of the parent hadron. $\widehat{{\cal P}}_{j}(x,b)$
is obtained from ${\cal P}_{j}(x,\vec{k}_{T},\zeta)$ by taking the
limit $\zeta\ra\infty$. The partial Sudakov factor ${\cal S}'(b,\zeta)$
in Eq.~(\ref{limzeta}) was derived in Ref.~\cite{Collins:1981uk}.
The function $\widehat{{\cal P}}_{j}(x,b)$ is the same in SIDIS and
Drell-Yan processes \nolinebreak\cite{Collins:2004nx}.

Among the three terms on the right-hand side of Eq.~(\ref{PHUP}),
only the hard vertex ${\cal H}_{j}^{(I)}$ depends on the sign of
the boson's virtuality. The functions ${\cal H}_{j}^{(T)}$ and ${\cal H}_{j}^{(S)}$
are given by the same Feynman graphs in the different crossed channels,
i.e., for bosons of timelike virtualities in the Drell-Yan case and
spacelike virtualities in the SIDIS case. Therefore, \begin{equation}
\overline{{\cal P}}_{j}^{(T)}(x,b)=\frac{\left|{\cal H}_{j}^{(T)}(b_{0}/b)\right|}{\left|{\cal H}_{j}^{(S)}(b_{0}/b)\right|}\overline{{\cal P}}_{j}^{(S)}(x,b)=r(b)\overline{{\cal P}}_{j}^{(S)}(x,b),\label{PHHP2}\end{equation}
where\begin{equation}
r(b)\equiv\frac{\left|{\cal H}_{j}^{(T)}(b_{0}/b)\right|}{\left|{\cal H}_{j}^{(S)}(b_{0}/b)\right|}=1+\alpha_{s}\left(\frac{b_{0}}{b}\right)\pi\frac{C_{F}}{4}+\dots\,.\label{HDYHSIDIS}\end{equation}
A similar consideration for $b=b_{*}$ proves the relationship (\ref{CinSCinT})
between ${\mathcal{C}}_{j/a}^{in\,(S)}(x,b_{*})$ and ${\mathcal{C}}_{j/a}^{in\,(T)}(x,b_{*})$,
as well as the relationship (\ref{cTcS}) between $c^{(T)}(x,b;b_{*})$
and $c^{(S)}(x,b;b_{*})$.

According to Eq.~(\ref{PHUP}), the dependence of the parton distributions
$\overline{{\cal P}}_{j}^{(T)}(x,b)$ and $\overline{{\cal P}}_{j}^{(S)}(x,b)$
on $x$ is determined by the same function $\widehat{{\cal P}}_{j}(x,b)$.
More generally, the $x$ dependence of the distributions $\overline{{\cal P}}_{a}^{(I)}(x,b)$
($a=g,u,d,...$) is determined by the type of the parent hadron and
parton flavor, but not by the type of the produced electroweak final
state. For example, the $x$ dependence of $\overline{{\cal P}}_{a}^{(I)}(x,b)$
is expected to be the same in the Drell-Yan process, $\gamma\gamma$
production via $q\bar{q}$ annihilation, and SIDIS; or in Higgs boson
production and $\gamma\gamma$ production via $gg$ fusion. This happens
because all dependence of $\overline{{\cal P}}_{a}^{(I)}(x,b)$ on
the hard-scattering process is determined by the hard vertex function
$|{\cal H}_{a}^{(I)}(b_{0}/b)|$ \cite{Catani:2000vq}, which does
not contain $x$. $\overline{{\cal P}}_{a}(x,b)$ are automatically
process-independent in the representations for $\widetilde{W}(b,Q,x_{A},x_{B})$
that always separate $\left|{\cal H}_{a}^{(I)}\right|$ from $\overline{{\cal P}}_{a}(x,b)$
\cite{Collins:1981uk,Collins:1982va,Catani:2000vq}. The best-fit
phenomenological functions ${\cal S}_{NP}^{BLNY}(b,Q;b_{*})$ and
${\cal S}_{NP}^{z}(b,Q;b_{*})$ were obtained in the CSS representation.
For this reason, we will use this representation in the numerical
analysis despite the additional complexities in the factorization
relations.

\section{\label{sec:Numerical-results}Numerical results}

\subsection{Overview\label{sub:Numerical-results-Overview}}

In this section, we compare the resummed cross sections (\ref{eq:full})
with the additional broadening term {[}$\rho(x)\neq0${]} to the resummed
cross sections without such a term {[}$\rho(x)=0${]}. We consider
the decay modes $W^{\pm}\rightarrow e\nu$, $Z^{0}\rightarrow e\bar{e},$
and $H^{0}\rightarrow\gamma\gamma$ and discuss the impact of the
experimental acceptance cuts imposed on the decay particles. The resummation
calculations that include the decay of the vector and scalar bosons
were described in Refs.~\cite{Balazs:1997xd} and \cite{Balazs:2000wv},
respectively. The perturbative Sudakov factor was included up to ${\mathcal{O}}(\alphas^{2})$,
and the functions $[{\mathcal{C}}^{in(T)}\otimes f]$ up to ${\mathcal{O}}(\alphas)$.
The numerical calculation was performed using the programs Legacy
and ResBos \cite{Landry:2002ix,Balazs:1997xd}, and with the CTEQ6M
parton distribution functions~\cite{Pumplin:2002vw}. 

\begin{figure}
\begin{center}\includegraphics[%
  clip,
  width=10cm,
  keepaspectratio]{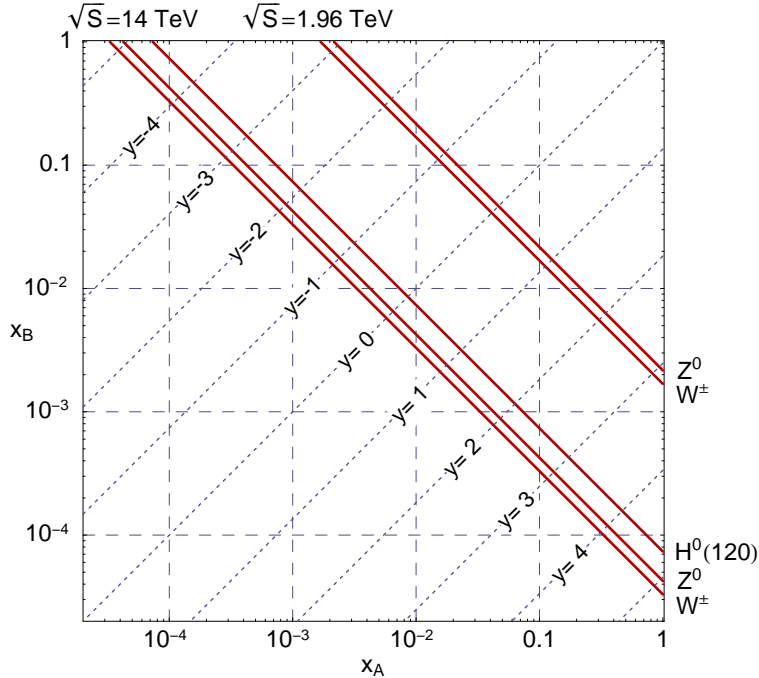}\vspace{-17pt}\end{center}

\caption{\label{fig:x1x2} Parton momentum fractions $x_{A}$ and $x_{B}$
accessible in $W,$ $Z,$ and Higgs boson production (for $M_{H}=120$
GeV) in the Tevatron Run-2 ($\sqrt{S}=1.96$ TeV) and at the LHC ($\sqrt{S}=14$
TeV). The accessible ranges of $x_{A}$ and $x_{B}$ are shown by
the solid lines. The contours of the constant rapidity $y$ are shown
by the inclined dotted lines. }
\end{figure}

\begin{figure}
\includegraphics[%
  width=0.32\textwidth,
  keepaspectratio]{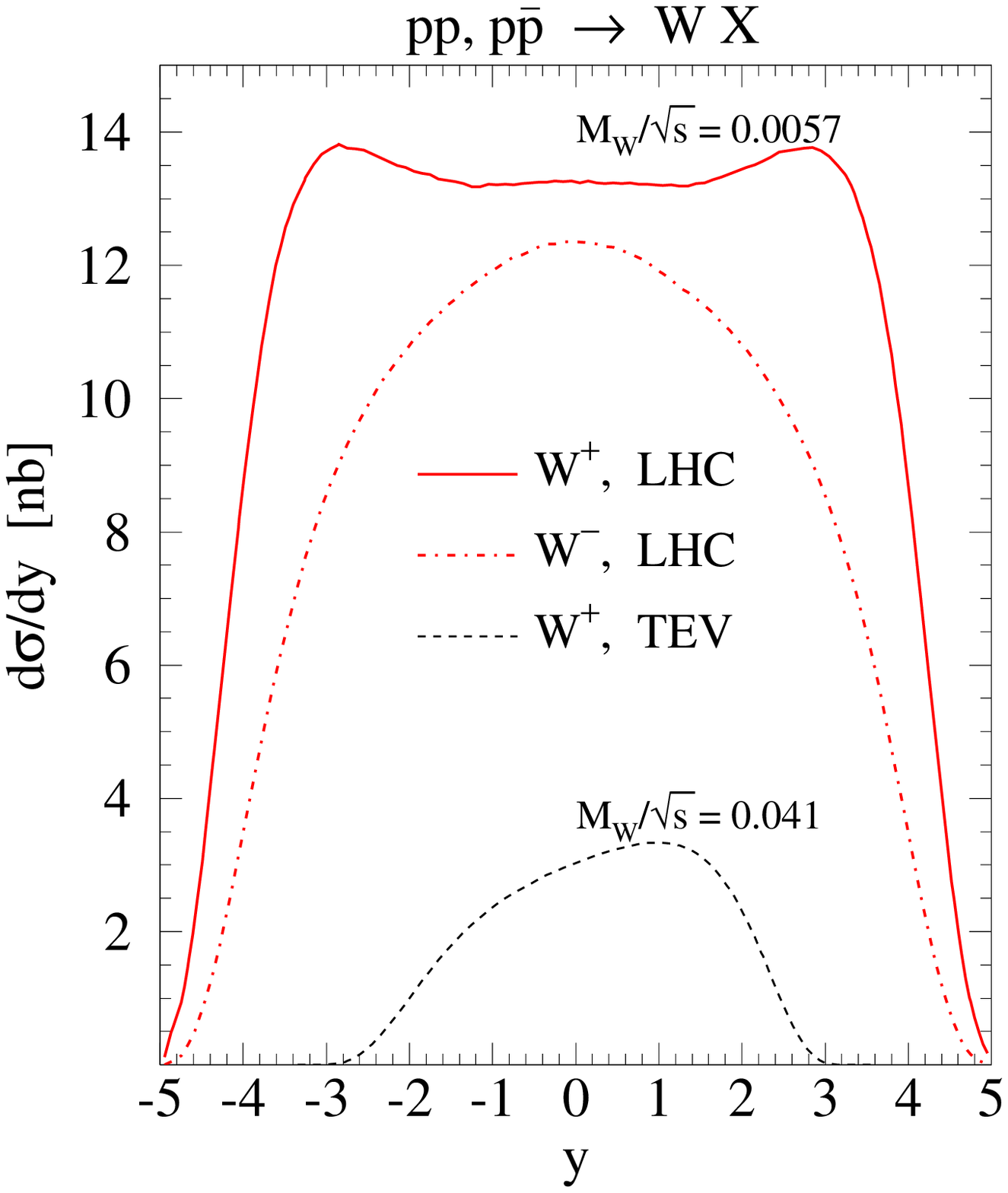} \includegraphics[%
  width=0.32\textwidth,
  keepaspectratio]{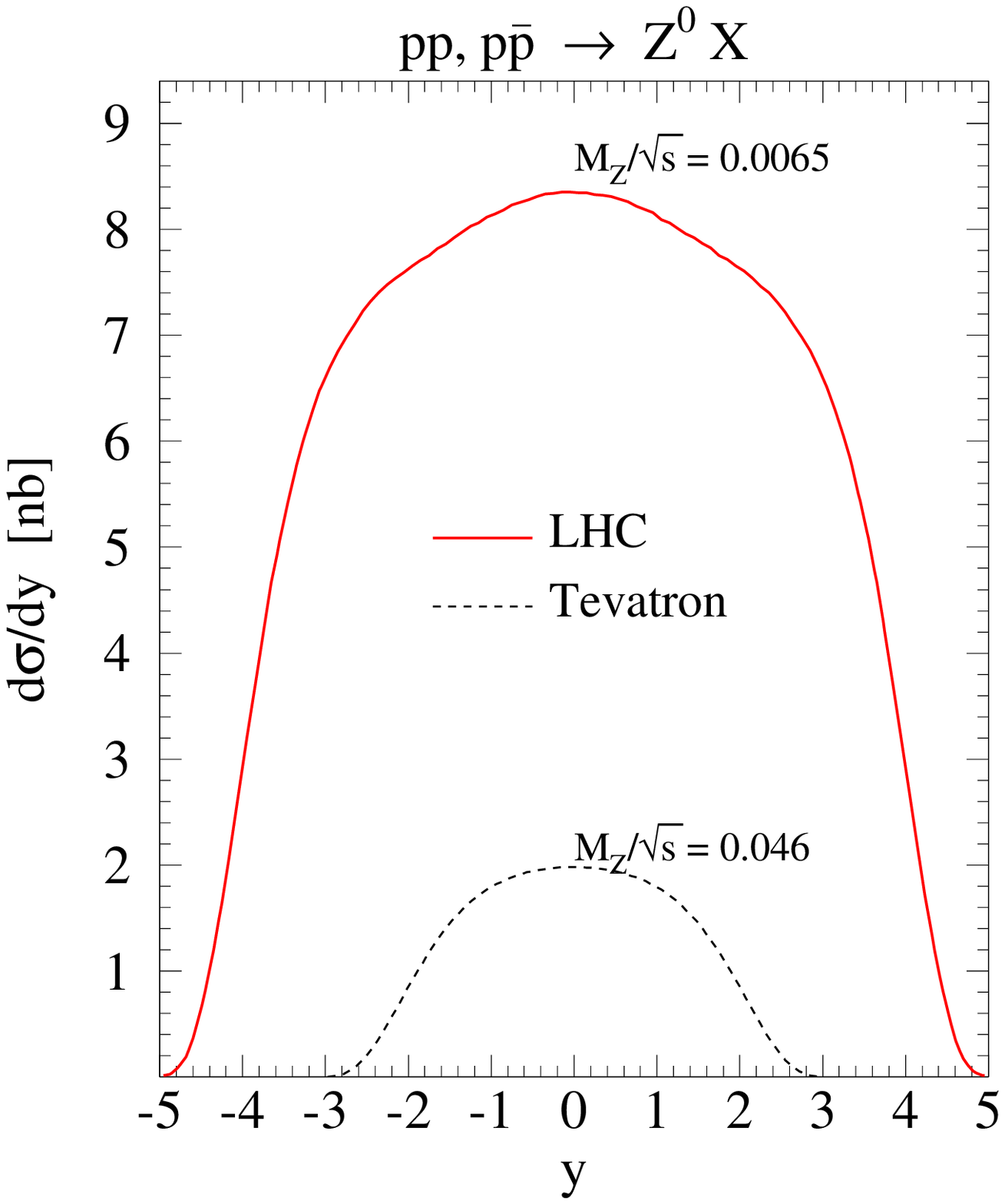} \includegraphics[%
  width=0.32\textwidth,
  keepaspectratio]{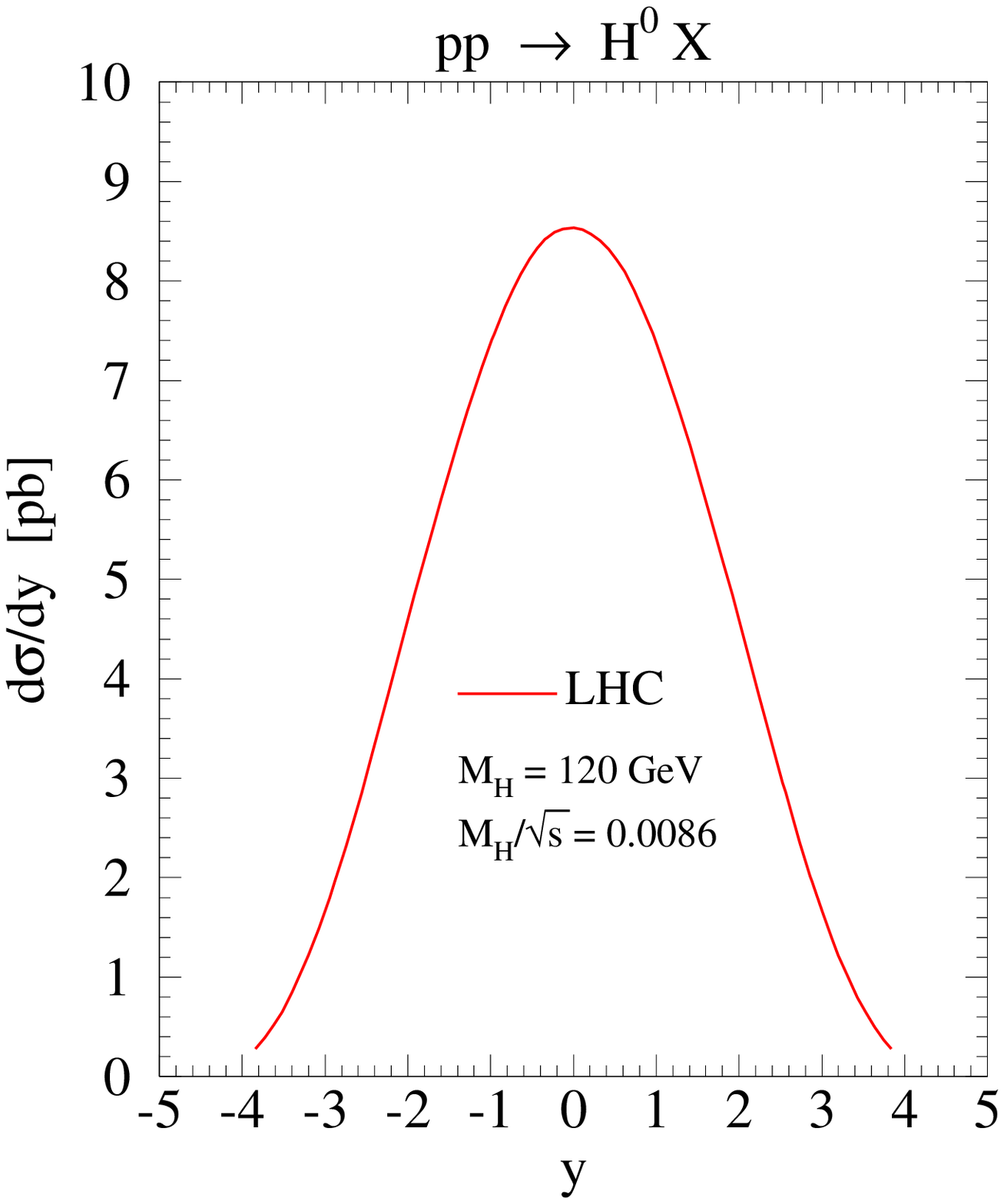}\vspace*{-10pt}\\

\hspace{.55cm}(a)\hspace{4.91cm}(b)\hspace{4.91cm}(c)\vspace*{-9pt}\\

\caption{Rapidity distributions $d\sigma/dy$ for (a) $W^{\pm}$ boson production
at the Tevatron and LHC; (b) $Z^{0}$ boson production at the Tevatron
and LHC; and (c) Higgs boson production at the LHC.\label{fig:rap} }
\end{figure}

For the chosen parameters, the small-$x$ broadening occurs when one
or both longitudinal momentum fractions $x_{A,B}\approx M_{V}e^{\pm y}/\sqrt{S}$
are of order, or less than, $x_{0}=0.005$. The accessible ranges
of $x_{A},$ $x_{B}$, and $y$ at the Tevatron and LHC are displayed
in Fig.~\ref{fig:x1x2}. Lower values of $x_{A}$ can be reached
at the price of pushing $x_{B}$ closer to unity, and vice versa.
The number of events affected by the broadening can be determined
from the rapidity distributions $d\sigma/dy$ in Fig.~\ref{fig:rap}.%
\footnote{The rapidity distributions and other observables inclusive in $q_{T}$
are insensitive to the $q_{T}$ broadening in our model, due to the
Gaussian form of the broadening exponential assumed in Eq.~(\ref{eq:full}).%
} In all scattering processes, most of the events occur at relatively
small $\left|y\right|$. The rate in the forward regions is suppressed
by the decreasing parton densities at $x\rightarrow1$. The broadening
can be detected at all boson rapidities if $\sqrt{S}\gtrsim M_{V}/x_{0}$,
or at forward rapidities $\left|y\right|\gtrsim\ln\left[M_{V}/(\sqrt{S}x_{0})\right]$,
if the cross section is large enough to measure the $q_{T}$ distribution
at such $\left|y\right|$. The heavy bosons are identified in the
experiment by observation of secondary particles from their decay.
The magnitude of the broadening depends on the procedure applied to
select the decay leptons or photons. The estimates for $W$ bosons
are modified by the integration over the unobserved rapidity of the
neutrino, which mixes contributions from different ranges of $y$.
The impact of the boson's decay will be addressed quantitatively in
the next subsections.

In the Tevatron Run-2 ($\sqrt{S}$=1.96 TeV), the $W$ and $Z$ bosons
are predominantly produced at $x_{A}\sim x_{B}\sim M_{V}/\sqrt{S}>x_{0}$.
The broadening can only affect a relatively small fraction of the
$W$ and $Z$ bosons with $\left|y\right|\gtrsim\ln\left[M_{V}/(\sqrt{S}x_{0})\right]\sim2$,
and it has negligible influence on most of the Tevatron observables.
The feasibility of discovering $q_{T}$ broadening at the Tevatron
depends on the available detector acceptance in the forward rapidity
regions. The rapidity coverage of the \D0 and CDF detectors is reviewed
in the appendix. Our estimates suggest that verification of $q_{T}$
broadening at the Tevatron is viable in the near future.

At the LHC ($\sqrt{S}$=14 TeV), at least one momentum fraction $x_{A}$
or $x_{B}$ is of order, or less than, $x_{0}$ in \emph{}all \emph{}$W$
and $Z$ boson events. Consequently the broadening is important at
all rapidities. The magnitude of the broadening at the LHC generally
differs between $Z$, $W^{+}$, and $W^{-}$ bosons, due to the differences
in the bosons' masses and rapidity distributions. The smaller mass
of $W$ bosons is conducive to the broadening. $70\%$ of $W^{+}$
bosons are produced in $u\bar{d}$ annihilation, enhanced at $\left|y\right|>1.5$
by contributions from valence $u$ quarks. Most of $W^{-}$ and $Z$
bosons are produced in scattering of valence $d$ quarks and/or sea
quarks, which tends to put more events in the central rapidity region.
Consequently the fraction of forward rapidity events is larger in
$W^{+}$ boson production than in $W^{-}$ or $Z$ boson production,
and, of the three boson species, the $W^{+}$ bosons are the most
sensitive to the broadening. The decay of $W^{\pm}$ bosons mixes
contributions from different ranges of $\left|y\right|$, reducing
the differences between $q_{T}$ broadening effects in the lepton-level
observables.

The Higgs bosons with the mass in the experimentally allowed range
($M_{H}$> 115 GeV) are produced via gluon fusion at somewhat larger
momentum fractions, such as $M_{H}/\sqrt{S}\sim0.0086$ for $M_{H}=\nolinebreak120$~GeV.
Their rapidity distribution in Fig.~\ref{fig:rap}(c) is narrower
than in the $W$ and $Z$ boson cases due to the suppression of the
forward rapidity regions by the decreasing gluon density $g(x)$ at
$x\rightarrow1$. Consequently the broadening is reduced in the Higgs
boson signal, but may be important in the background processes.

Next, we present the quantitative results for the individual scattering
processes.

\subsection{$W$ and $Z$ boson production in the Tevatron Run-2\label{sub:Tev2-W-Z}}

\subsubsection{$Z$ boson production}

\begin{figure}[p]
\begin{center}\includegraphics[%
  scale=0.8]{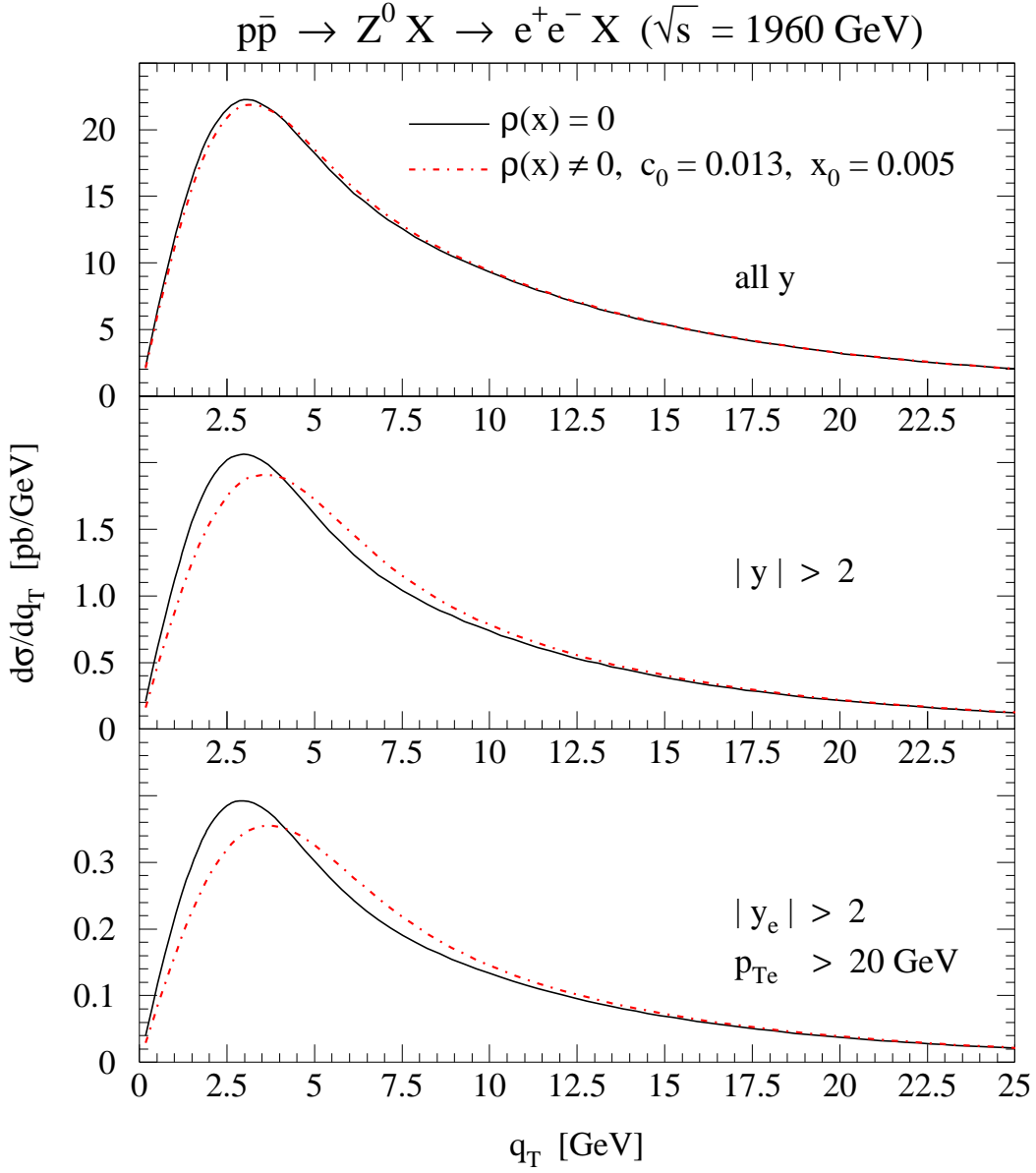}\vspace*{-17pt}\end{center}

\caption{Transverse momentum distributions of $Z$ bosons in the Tevatron
Run-2: (a) integrated over the full range of $Z$ boson rapidity $y$;
(b)~for events with a large $Z$ boson rapidity, $|\, y\,|>2$; (c)
for events with both decay electrons registered in the forward ($y_{e^{+}}>2$,
$y_{e^{-}}>2$, and $p_{Te^{\pm}}>20$ GeV) or backward ($y_{e^{+}}<-2$,
$y_{e^{-}}<-2$, and $p_{Te^{\pm}}>20$~ GeV) detector regions. The
solid curve is the standard CSS cross section, calculated using the
BLNY parametrization~\cite{Landry:2002ix} of the nonperturbative
Sudakov factor. The dashed curve includes additional terms responsible
for the $q_{T}$ broadening in the small-$x$ region, as in Eq.~(\ref{eq:full}).\label{fig:zTeV}}
\end{figure}

The broadening may be most easily observed in the dilepton channel
in $Z$ boson production in the Tevatron Run-2. The strategy here
is to exclude contributions from the central-rapidity $Z$ bosons,
which are almost insensitive to the broadening. If no distinction
between the central and forward $Z$ bosons is made (as, e.g., in
the Run-1 analysis), the small-$x$ broadening contributes at the
level of the other uncertainties in the resummed form factor. Fig.~\ref{fig:zTeV}(a)
shows the $Z$ boson distribution $d\sigma/dq_{T}$, integrated over
the $Z$ boson rapidity $y$ without selection cuts on the decay leptons.
The cross section with the broadening term (dashed line) essentially
coincides with the cross section without such a term (solid line).
Both cross sections are dominated by contributions from $x\sim M_{Z}/\sqrt{S}\sim0.046\gg x_{0}$,
where the broadening function $\rho(x)$ is negligible. 

In contrast, the small-$x$ broadening visibly modifies $d\sigma/dq_{T}$
at forward rapidities, where one of the initial-state partons carries
a smaller momentum fraction than in the central region. Fig.~\ref{fig:zTeV}(b)
shows the cross section $d\sigma/dq_{T}$ for $Z$ bosons satisfying
$|y|>2$. The peak of the curve with $\rho(x)\neq0$ is lower and
shifted toward higher $q_{T}$. Fig.~\ref{fig:zTeV}(c) displays
the cross sections with the acceptance cuts $y_{e^{\pm}}\,>2,$ $p_{Te^{\pm}}>20$
GeV or $y_{e^{\pm}}\,<-2,$ $p_{Te^{\pm}}>20$ GeV simultaneously
imposed on both decay leptons. The cuts exclude contributions with
$\left|y\right|<2$ and retain a fairly large cross section ($\approx3.4$
pb), most of which falls within the acceptance region (extending up
to $\left|y_{e}\right|\approx3$ in the \D0 detector). The Run-2
can probably discriminate between the curves in Fig.~\ref{fig:zTeV}(c)
given the improved acceptance and increased luminosity of the upgraded
Tevatron collider.

To estimate sensitivity of the Tevatron $Z$ boson production to uncertainties
in the model for $q_{T}$ broadening, we vary the control parameter
$x_{0}$ {[}cf.~Eq.~(\ref{eq:ax}){]} within the limits compatible
with the analyzed SIDIS $z$-flow data. The parameter $x_{0}$ determines
the upper boundary of the $x$ range in which the broadening effects
are important. As was observed in Ref.~\cite{Nadolsky:2000ky}, the
$q_{T}$-broadening was required in most of the range $x\in[10^{-4},\,10^{-2}]$
in SIDIS in order to obtain a good fit to $d\Sigma_{z}/dq_{T}$ from
Refs.~\cite{Aid:1995we,Adloff:1999ws}. Given this observation, $x_{0}=0.005$
assumed in Fig.~\ref{fig:zTeV} is a conservative estimate of the
upper boundary of the $x$ region impacted by the broadening at HERA.
Values of $x_{0}$ as high as $10^{-2}$ are compatible with the HERA
data, while $x_{0}\sim10^{-3}$ or smaller would lead to a poor fit
to the HERA data in the range $x\in[x_{0},\,10^{-2}]$. 

\begin{figure}
\includegraphics[%
  scale=0.8]{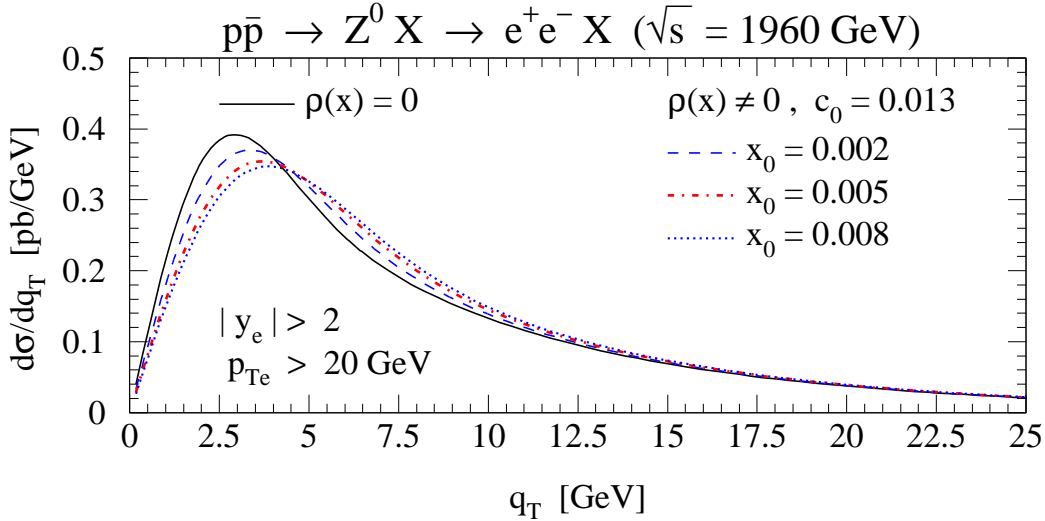}

\caption{Same as Fig.~\ref{fig:zTeV}(c), for the BLNY parametrization (solid),
and various choices of the parameter $x_{0}$ in Eq.~(\ref{eq:ax}):
$x_{0}=0.002$ (dashed), $0.005$ (dot-dashed), and $0.008$ (dotted).\label{fig:uncert}}
\end{figure}

If $x_{0}$ lies in the range $10^{-3}-10^{-2}$, the broadening remains
observable at the Tevatron. Fig.~\ref{fig:uncert} shows the cross
sections for $x_{0}=\{0.002,0.005,0.008\}$, compared to the standard
BLNY result. Although the broadening is reduced for $x_{0}=0.002,$
it may still be discernible in a high-statistics data sample. On the
other hand, the broadening can be also enhanced for $x_{0}>0.005,$
as demonstrated by the curve for $x_{0}=0.008.$ The magnitude of
the broadening for a given $x_{0}$ is affected by the parameter $c_{0}$,
which is anti-correlated with the power of $x$ in the phenomenological
factor $e^{-c_{0}b^{2}/x^{p}}$ found in the SIDIS fit. Adjustments
in $c_{0}$ can be compensated in some range by varying $p$ and without
deteriorating the overall quality of the SIDIS fit. For example, the
central value $c_{0}=0.013$ was obtained for $p=1$ in the best fit
\cite{Nadolsky:2000ky} to both sets of the $z$-flow data \cite{Aid:1995we,Adloff:1999ws}.
The best fit to the earlier of the two H1 data sets \cite{Aid:1995we}
was obtained in Ref.~\cite{Nadolsky:1999kb} for $e^{-{\cal S}_{NP}^{z}}=\exp\left[-4.58+0.58/\sqrt{x}\right]$,
\emph{i.e.}, reduction in the negative power of $x$ from $1$ to
0.5 was partly compensated by the increase of $c_{0}$ from 0.013
to 0.58. The magnitude of the broadening at the Tevatron for the superseded
$1/\sqrt{x}$ parametrization is even larger than that shown in Fig.~\ref{fig:uncert}.
Correlated uncertainties in the parameters of the SIDIS fit (and possibly
a combined SIDIS/Drell-Yan fit) can be constrained in a future dedicated
study, especially after improved SIDIS data from HERA-2 become available.

\subsubsection{$W$ boson production}

\begin{figure}
\begin{center}\includegraphics[%
  clip,
  scale=0.8]{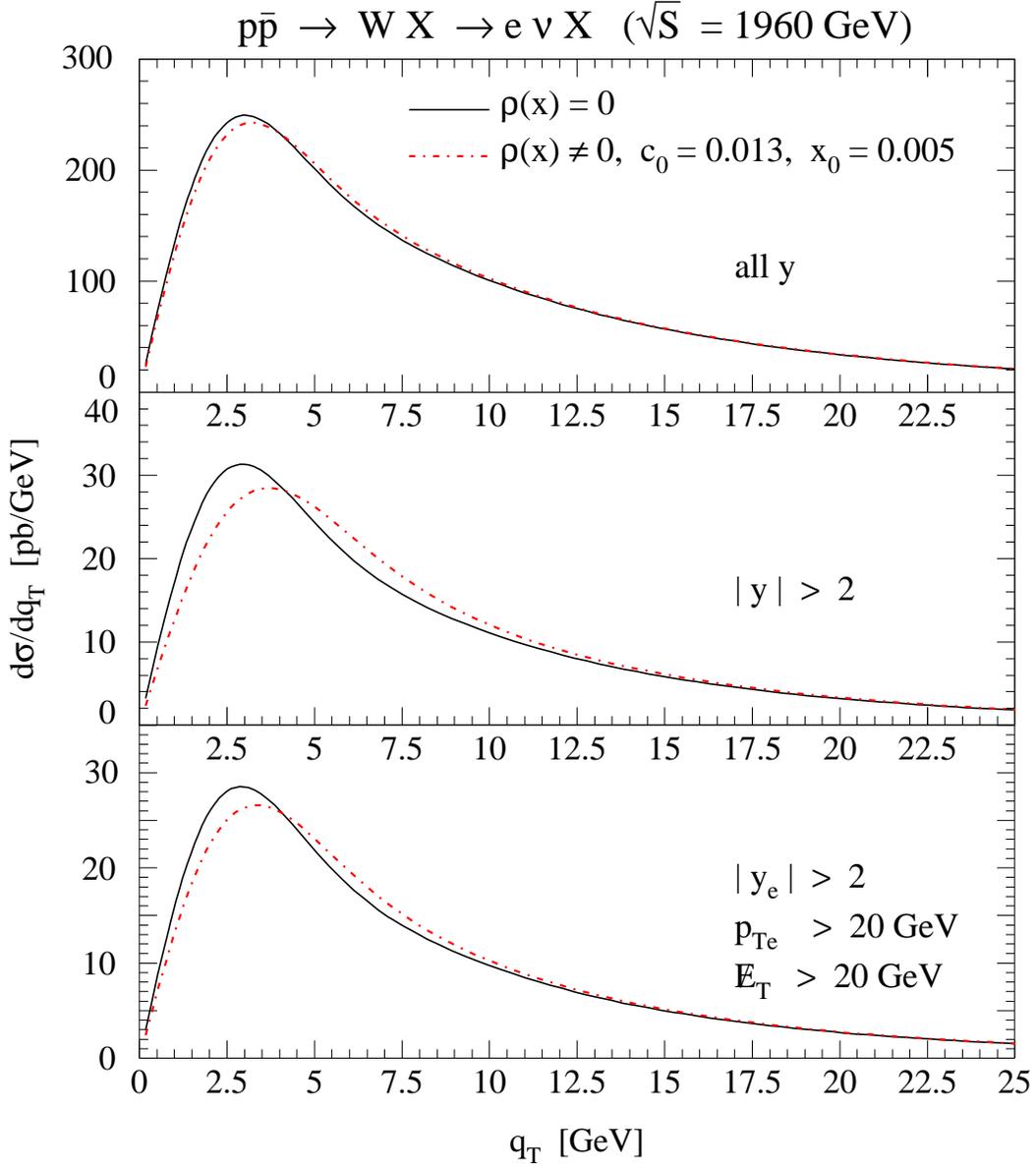}\vspace*{-17pt}\end{center}

\caption{Transverse momentum distributions for a combined sample of $W^{+}$
and $W^{-}$ bosons in the Tevatron Run-2; (a) integrated over the
full range of $W$ boson rapidities; (b)~integrated over the forward
boson rapidities $|\, y\,|>2$; (c) integrated with the selection
cuts $|\, y_{e}\,|>2,\,$ $p_{Te}>20$ GeV, $\, E_{T}\hspace{-13pt}/\hspace{8pt}>20\,$GeV.
The solid curve is the standard CSS cross section, calculated using
the BLNY parametrization~\cite{Landry:2002ix} of the nonperturbative
Sudakov factor. The dashed curve includes additional terms responsible
for the $q_{T}$ broadening in the small-$x$ region, as in Eq.~(\ref{eq:full}).\label{fig:wTeV}}
\end{figure}

In the absence of leptonic cuts, $W$ boson production is slightly
more sensitive to the small-$x$ broadening as compared to $Z$ boson
production because of the smaller mass and wider rapidity distribution
of the $W$ boson. The relevant $q_{T}$ distributions are displayed
in Fig.~\ref{fig:wTeV}. As in the case of the $Z$ boson, the broadening
is essentially negligible in the absence of selection cuts on $y$
or leptonic momenta {[}cf.~Fig.~\ref{fig:wTeV}(a){]}. In the forward
region $|y|>2$ {[}Fig.~\ref{fig:wTeV}(b){]}, the broadening term
shifts the peak of $d\sigma/dq_{T}$ significantly. The change $\delta q_{T}\approx\nolinebreak750$~MeV
in the position of the peak is larger than the analogous change $\delta q_{T}\approx\nolinebreak500$~MeV
in the $Z$ boson cross section. 

Integration over the unobserved neutrino's rapidity in the $e\nu$
decay channel mixes contributions from different rapidities of $W$
bosons, thus enhancing the broadening at central electron rapidities
and reducing the broadening at forward electron rapidities. The strongest
effect occurs when only forward electrons are selected. Fig.~\ref{fig:wTeV}(c)
shows the cross sections $d\sigma/dq_{T}$ with constraints imposed
on the momentum of the electron ($|y_{e}|>2,$ $p_{Te}>20$ GeV) and
missing transverse energy associated with the neutrino ($\, E_{T}\hspace{-13pt}/\hspace{10pt}>20$
GeV). Although visible, the difference between the solid and dashed
curves in Fig.~\ref{fig:wTeV}(c) is less pronounced than in the
$W$ boson-level distribution in Fig.~\ref{fig:wTeV}(b) or in the
$Z$-boson lepton-level distribution in Fig.~\ref{fig:zTeV}(c).
This is because some events in the sample with a forward electron
rapidity and arbitrary neutrino rapidity are due to the decay of central-rapidity
$W$ bosons, which are not affected by the broadening. In contrast,
events with two forward high-$p_{T}$ electrons in $Z$ boson production
are always due to the decay of forward-boosted $Z$ bosons, and they
are more sensitive to the broadening effects.

Despite its modest magnitude, the forward $q_{T}$ broadening may
affect the precision measurement of the $W$ boson mass $M_{W}$ in
the Tevatron Run-2. The current goal of CDF and \D0 experiments
is to measure $M_{W}$ with the accuracy of 30-40 MeV per experiment.
The $W$ boson mass is commonly extracted from the distribution with
respect to the leptonic transverse mass \cite{Smith:1983aa}, defined
by \begin{equation}
M_{T}^{e\nu}\equiv\sqrt{\left(\left|\vec{p}_{Te}\right|+\left|\vec{p}_{T\nu}\right|\right)^{2}-\left(\vec{p}_{Te}+\vec{p}_{T\nu}\right)^{2}}\label{eq:MT}\end{equation}
in terms of the transverse momentum $\vec{p}_{Te}$ of the electron
and transverse momentum $\vec{p}_{T\nu}$ of the neutrino. In the
experiment, the neutrino's transverse momentum is equated to the missing
transverse energy, $\left|\vec{p}_{T\nu}\right|=E_{T}\hspace{-13pt}/\hspace{8pt}$.
The determination of $M_{W}$ from the shape of the kinematical (Jacobian)
peak in $d\sigma/dM_{T}^{e\nu}$ is scarcely sensitive to the boson's
transverse momentum $q_{T}$.%
\footnote{More precisely, a variation $\delta q_{T}$ in the typical transverse
momentum results in an error of order $\delta M_{T}^{e\nu}/M_{T}^{e\nu}\sim\delta q_{T}^{2}/M_{W}^{2}\ll1$
in the typical transverse mass. %
} We have verified that the impact of the broadening on $d\sigma/dM_{T}^{e\nu}$
is negligible. However, reconstruction of $E_{T}\hspace{-13pt}/\hspace{8pt}$
introduces a sizable uncertainty in the determination of $M_{W}$.
An alternative method is to extract $M_{W}$ from the distribution
$d\sigma/dp_{Te}$. This method does not suffer from the complications
associated with the reconstruction of $E_{T}\hspace{-13pt}/\hspace{8pt}$,
but, as a trade-off, it is directly sensitive to the transverse momentum
$q_{T}$ of the $W$ bosons. In practice, the final uncertainties
on $M_{W}$ from the transverse mass and transverse momentum methods
are comparable. As we will now show, the $q_{T}$ broadening can substantially
affect $M_{W}$ determined from the transverse momentum of electrons
in the forward calorimeter. 

\begin{figure}
\begin{center}\includegraphics[%
  width=9cm,
  keepaspectratio]{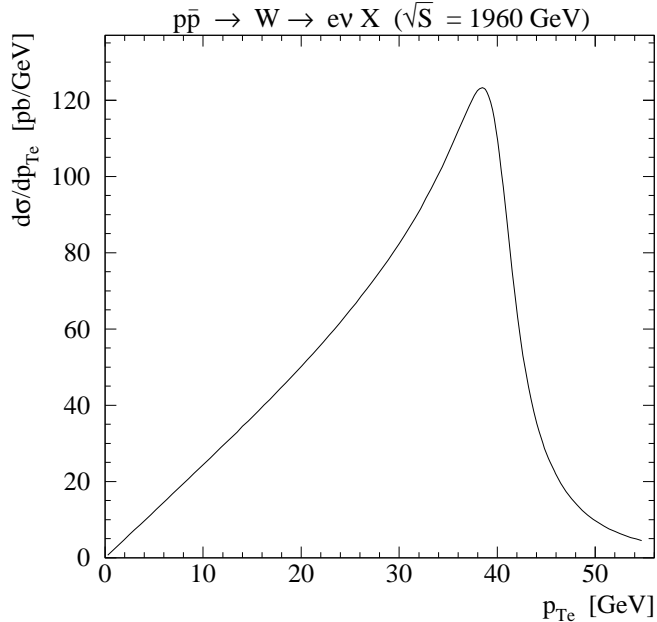}\vspace{-17pt}\end{center}

\caption{\label{fig:W-Tev-pTe}Transverse momentum distribution $d\sigma/dp_{Te}$
of the electrons from the decay of $W$ bosons in the Tevatron Run-2.
CTEQ6M parton densities \cite{Pumplin:2002vw} and BLNY nonperturbative
Sudakov factor \cite{Landry:2002ix} were used.}
\end{figure}

\begin{figure}
\begin{center}\includegraphics[%
  width=0.49\textwidth,
  keepaspectratio]{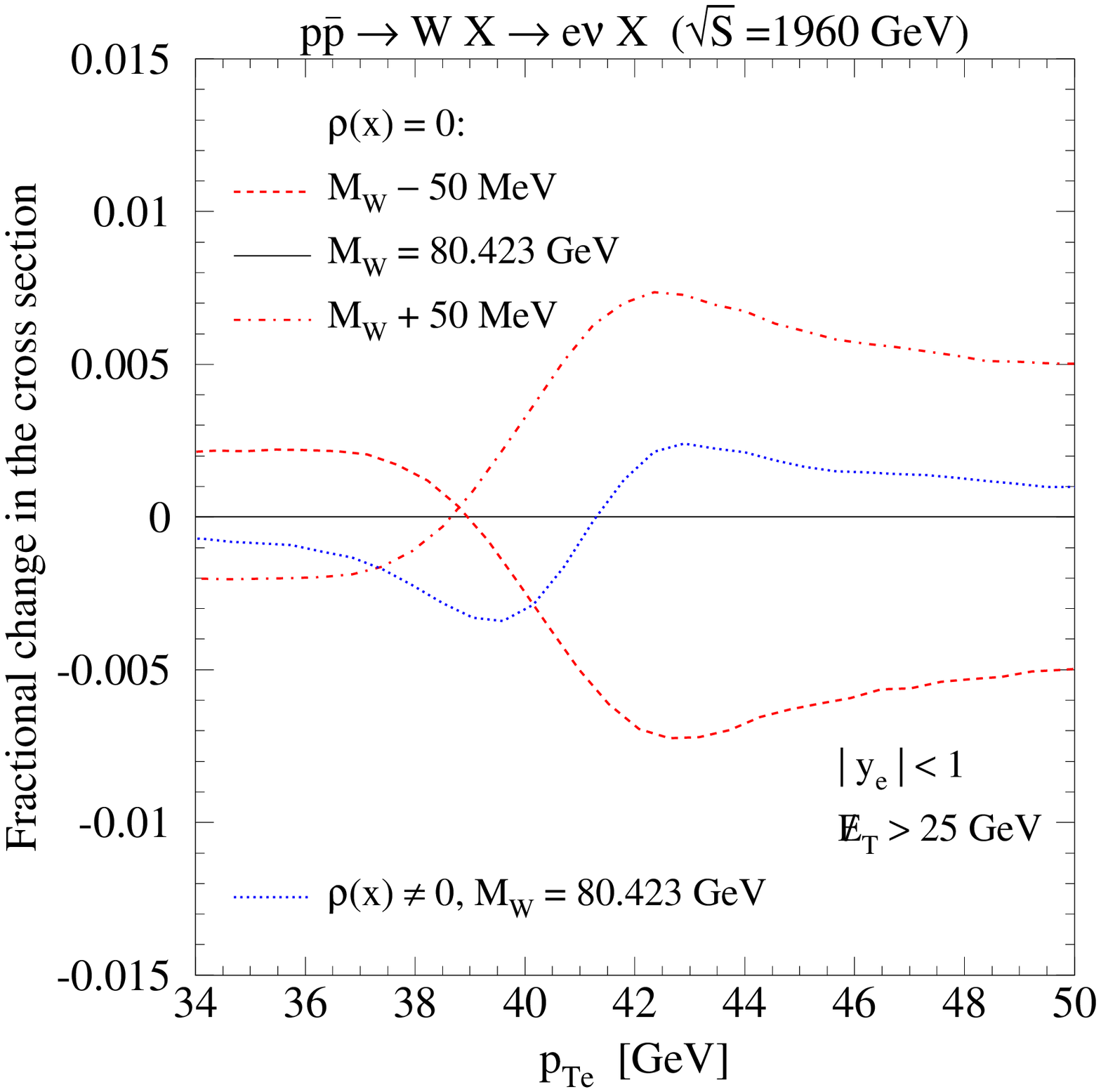} \includegraphics[%
  width=0.49\textwidth,
  keepaspectratio]{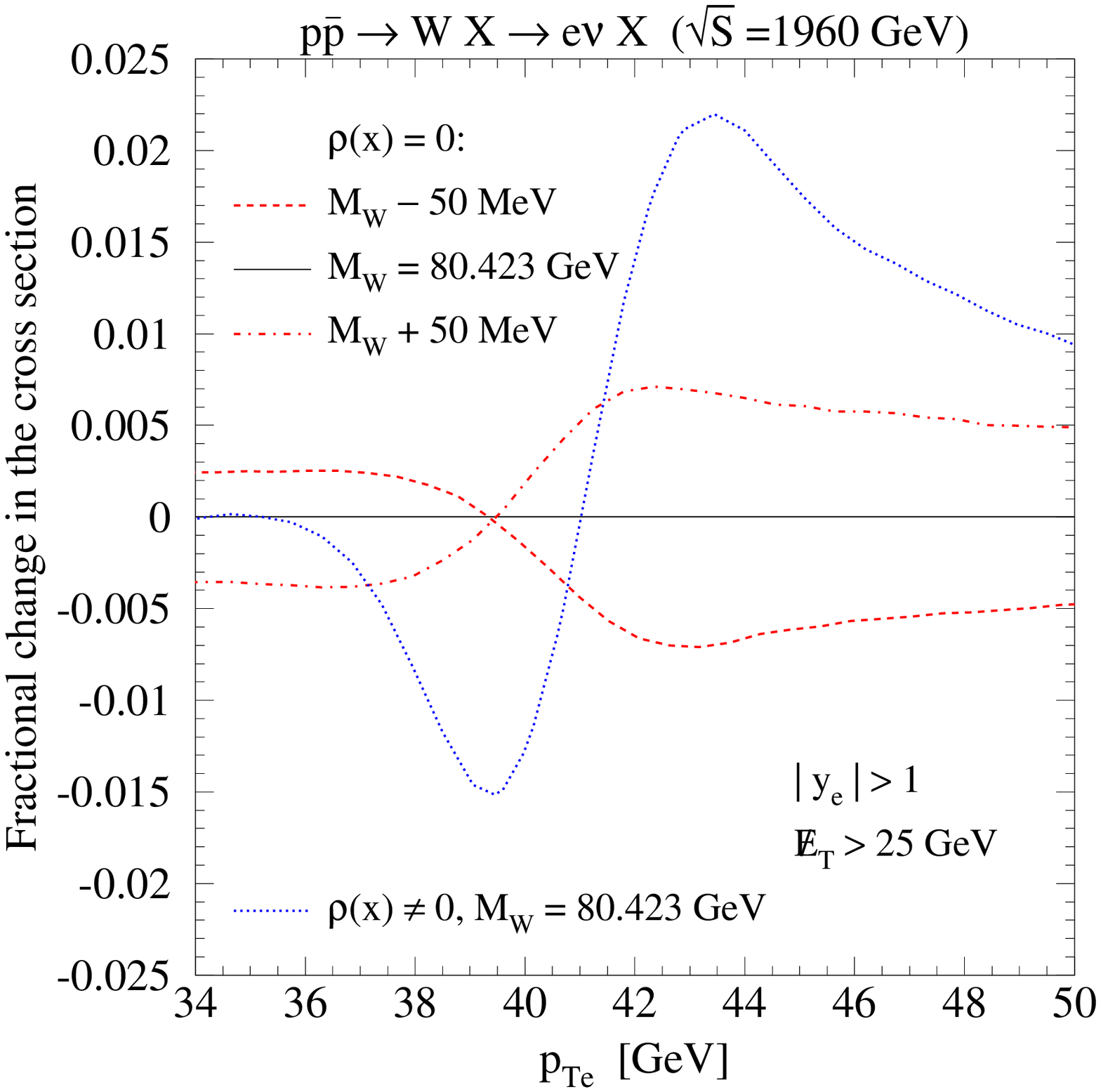}\\
\hspace{1.2cm} (a)\hspace{7.7cm}(b)\vspace*{-17pt}\\ \end{center}

\caption{The fractional difference in the distribution $d\sigma/dp_{Te}$:
(a) for the central-rapidity sample of electrons ($|y_{e}|<1$), and
(b) for the forward-rapidity sample of electrons ($|y_{e}|>1$). The
shown curves are explained in the main text. \label{fig:wPTratio}}
\end{figure}

The distribution $d\sigma/dp_{Te}$ is shown in Fig.~\ref{fig:W-Tev-pTe}.
Its Jacobian peak is located at $p_{Te}\sim\nolinebreak M_{W}/2\approx40$
GeV. To better visualize percent-level changes  in $d\sigma/dp_{Te}$
associated with the broadening, we plot in Fig.~\ref{fig:wPTratio}
the fractional difference $\left(d\sigma^{mod}/dp_{Te}\right)/\left(d\sigma^{std}/dp_{Te}\right)-1$
of the {}``modified'' (\emph{mod}) and {}``standard'' (\emph{std})
theory cross sections. The standard cross sections are obtained by
taking $\rho(x)=0$ and some reference value of $M_{W}$, which we
choose to be equal to $80.423$~GeV. The modified cross sections
are obtained by taking either $\rho(x)\neq0$ (dotted curve) or a
slightly varied $W$ boson mass, $M_{W}+50$ MeV (dashed curve) and
$M_{W}-50$ MeV (dot-dashed curve). The impact of $q_{T}$ broadening
on the measurement of $M_{W}$ can be judged by comparing the fractional
differences due to the inclusion of the broadening term $\rho(x)\neq0$
and explicit variation of $M_{W}$ by $\pm50$ MeV for $\rho(x)=0$.

A small increase of $M_{W}$ results in a positive shift of the Jacobian
peak, which reduces the cross section at $p_{Te}<M_{W}/2$ and increases
the cross section at $p_{Te}>M_{W}/2.$ A decrease in $M_{W}$ shifts
the Jacobian peak in the opposite direction. The broadening of $d\sigma/dq_{T}$
shifts the Jacobian peak in the positive direction. At $|y_{e}|<1$
or in the rapidity-inclusive sample, the broadening can be mistaken
for an increase of $M_{W}$ by 10-20 MeV, as well as uncertainties
in the parton distributions and/or nonperturbative Sudakov factor
{[}cf. Fig.~\ref{fig:wPTratio}(a){]}. At $|y_{e}|>1$, the small-$x$
broadening exceeds the other theoretical uncertainties and is comparable
with a variation of $M_{W}$ by more than 50~MeV {[}cf. Fig.~\ref{fig:wPTratio}(b){]}.
It remains observable even if the electron's energy is smeared due
to the finite resolution of the detector. The experimental fractional
resolution $\delta E/E$ near the Jacobian peak is typically less
than $3\%$ \cite{Affolder:2000bp,Abazov:2002bu}. We have estimated
the energy smearing by convolving $d\sigma/dp_{Te}$ with the Gaussian
functions of the width corresponding to $\delta E/E$ between 1 and
5\%. In all cases, the variation in the smeared $d\sigma/dp_{Te}$
due to the broadening has exceeded variations due to the shift of
$M_{W}$ by $\pm50$ MeV.

\subsection{Electroweak boson production at the Large Hadron Collider\label{sub:LHC}}

\subsubsection{$W$ and $Z$ bosons\label{sub:LHC-W-Z}}

\begin{figure}
\begin{center}\includegraphics[%
  scale=0.8]{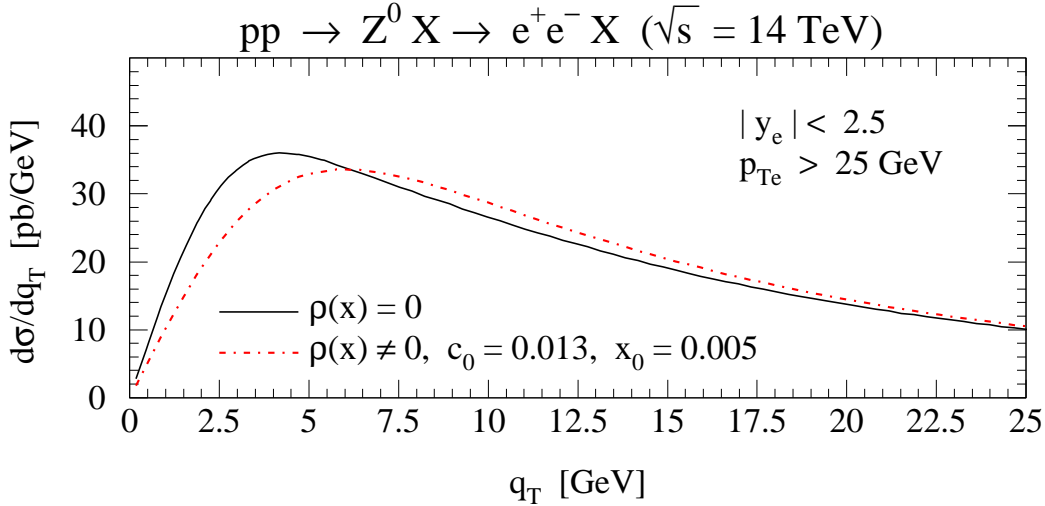}\vspace*{-17pt}\\\end{center}

\caption{\label{fig:ZLHC} Transverse momentum distributions of $Z$ bosons
at the Large Hadron Collider. The events are selected by requiring
$|\, y_{e}\,|<2.5$ and $p_{Te}>25$ GeV for both decay electrons.}
\end{figure}
\begin{figure}
\begin{center}\includegraphics[%
  scale=0.8]{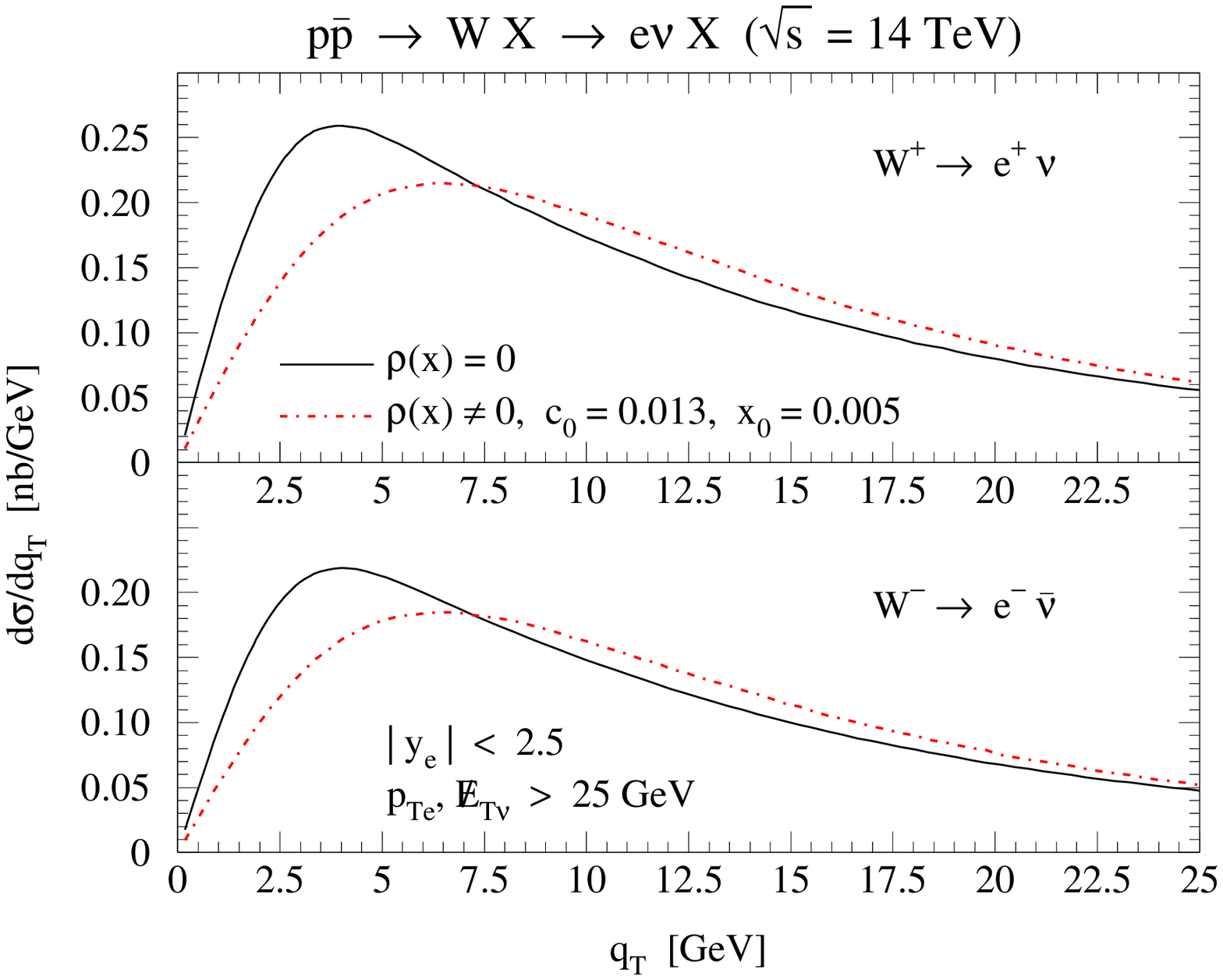}\vspace*{-17pt}\\\end{center}

\caption{\label{fig:WLHC} Transverse momentum distributions of (a) $W^{+}$
bosons and (b) $W^{-}$ bosons at the Large Hadron Collider. The decay
leptons are required to satisfy $|\, y_{e}\,|<2.5$, $\, p_{Te}>25$
GeV, $\, E_{T}\hspace{-13pt}/\hspace{13pt}>25\,$GeV. }
\end{figure}

At the LHC, the small-$x$ broadening may be observed in $W$ and
$Z$ boson production at all rapidities. Very forward vector bosons
probe $x$ down to $\sim3\cdot10^{-5}$, while the SIDIS data constrains
the $b$-dependent parton distributions at $x$ above $\sim10^{-4}$.
To stay in the kinematic region probed by the SIDIS data, we restrict
our analysis to the events in the central-rapidity region, which we
define by requiring the rapidities of the decay electrons to be less
than 2.5. 

Fig.~\ref{fig:ZLHC} demonstrates $q_{T}$ distributions of centrally
produced $Z$ bosons, selected by imposing the cuts $\left|y_{e}\right|<2.5$
and $p_{Te}>25$ GeV on both decay electrons. The distribution with
$\rho(x)\neq0$ is shifted toward higher $q_{T}$. The broadening
shift exceeds typical uncertainties in ${\cal S}_{NP}$: a similar
shift would result from the increase of ${\cal {\cal S}}_{NP}(b,M_{Z};b_{*})$
from $\approx3.2b^{2}$ in the BLNY parametrization to $\approx8b^{2}$.
The $q_{T}$ shift is even larger in production of $W$ bosons {[}cf.~Fig.~\ref{fig:WLHC}{]},
which we evaluate by requiring $|\, y_{e}|<2.5,\, p_{Te}>25$~GeV,
and $E_{T}\hspace{-13pt}/\hspace{8pt}>25$ GeV. The smaller mass $M_{W}$
and less restrictive leptonic cuts result in smaller typical $x$
probed by the $W$ bosons, hence, in stronger broadening. The shift
is slightly larger in $W^{+}$ boson production {[}cf.~Fig.~\ref{fig:WLHC}(a){]}
as compared in $W^{-}$ boson production {[}cf.~Fig.~\ref{fig:WLHC}(b){]}
because of the flatter $y$ distribution for $W^{+}$ bosons {[}cf.~Fig.~\ref{fig:rap}(a){]}.
 The shown $q_{T}$ broadening propagates into the leptonic transverse
mass and transverse momentum distributions. Both $M_{T}^{e\nu}$ and
$p_{T\ell}$ methods for the measurement of $M_{W}$ are affected
in this case, in contrast to the Tevatron, where the $M_{T}^{e\nu}$
method is not susceptible to the broadening.

\subsubsection{Higgs boson\label{sub:LHC-Higgs}}

\begin{figure}
\begin{center}\includegraphics[%
  width=0.48\textwidth,
  keepaspectratio]{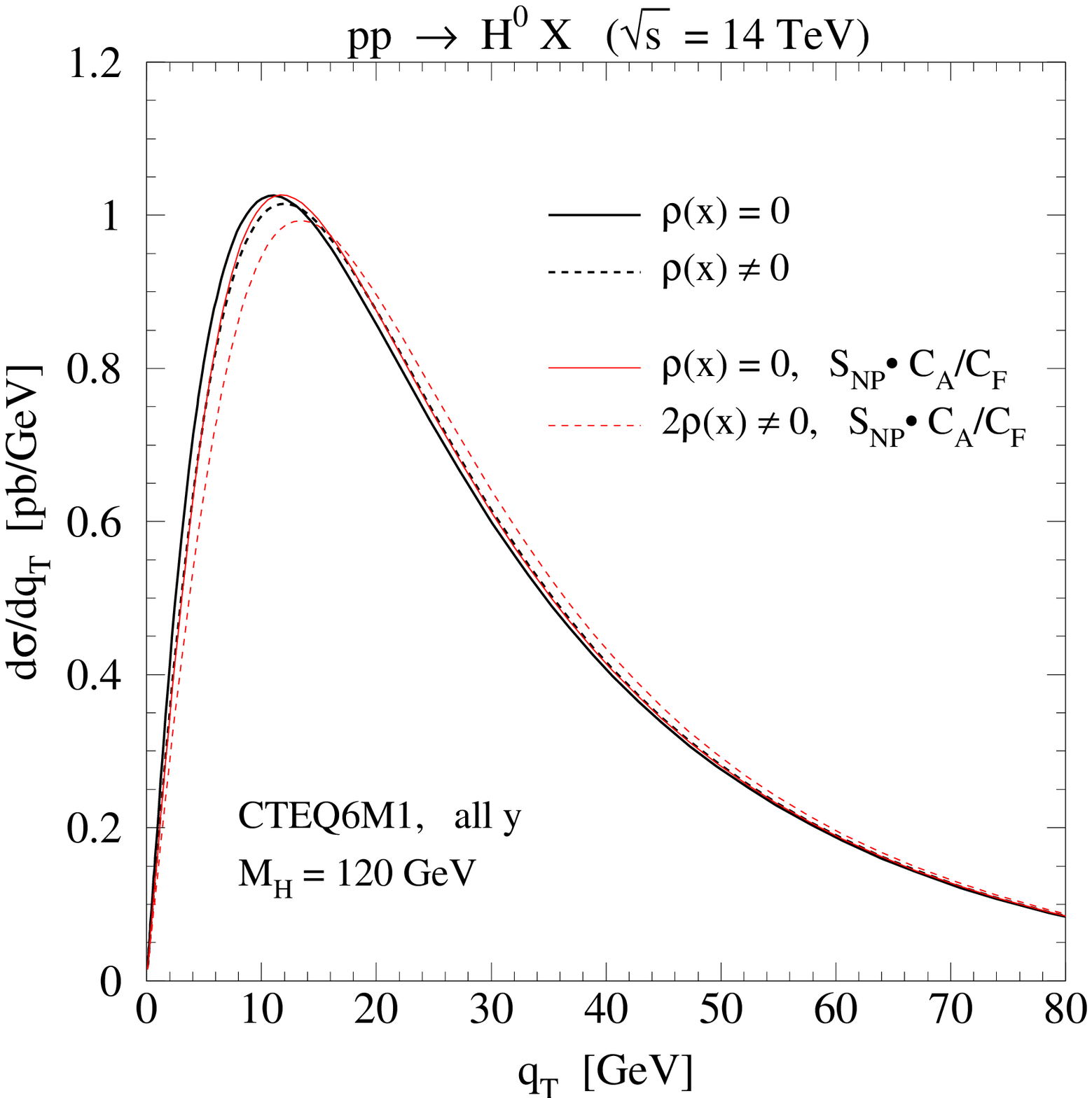} \includegraphics[%
  width=0.48\textwidth,
  keepaspectratio]{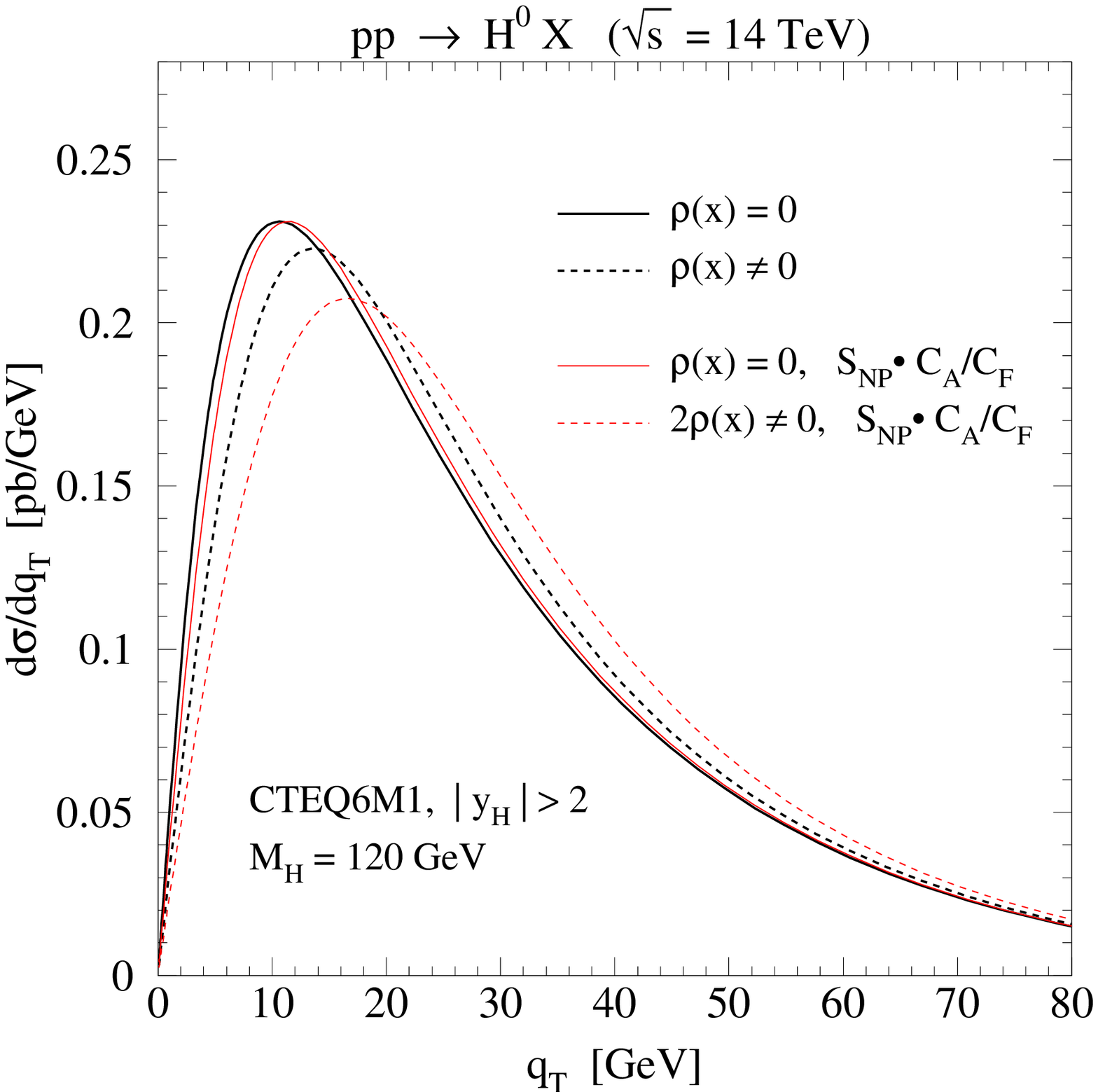}\\
\hspace{.2cm} (a)\hspace{7.7cm}(b)\vspace*{-20pt}\\\end{center}

\caption{The transverse momentum distribution $d\sigma/dq_{T}$ of on-shell
Higgs bosons at the Large Hadron Collider: (a) integrated over the
full range of boson rapidities; (b) integrated over the forward region
$|y|>2$. The meaning of the curves is explained in the text.\label{fig:HLHC}}
\end{figure}

The small-$x$ broadening is less spectacular, but visible, in the
production of light Higgs bosons via the effective $ggH$ vertex.
Standard model Higgs bosons with mass below about 140~GeV can be
distinguished above the continuum background by their decay into a
pair of highly energetic photons. The photon pairs produced from the
Higgs boson decay have a larger typical $q_{T}$ than the photon pairs
produced from the dominant background processes $q\bar{q}\rightarrow\gamma\gamma$
and $gg\rightarrow\gamma\gamma$.%
\footnote{We refer to each hard subprocess according to its Born-level contribution,
but the resummed cross sections include the relevant higher-order
corrections to the Born-level diagram. For instance, the resummed
cross section for $q\bar{q}\rightarrow\gamma\gamma$ also includes
the contributions from $q\bar{q}\rightarrow\gamma\gamma g,$ $qg\rightarrow\gamma\gamma q$,
etc. %
} For this reason, selection of photon pairs with $q_{T}$ above 20-30
GeV can improve the statistical significance of the Higgs boson signal.

The harder $q_{T}$ spectrum of the $gg$-dominated Higgs boson signal
results from the larger leading-logarithm coefficient ${\cal A}^{(1)}$
in the perturbative Sudakov factor (equal to $C_{A}=3$ in $gg$ channels
and $C_{F}=4/3=(4/9)C_{A}$ in $q\bar{q}$ channels). The Higgs boson
$q_{T}$ distribution peaks at $q_{T}=10-20$ GeV, as compared to
$q_{T}\sim5-10$ GeV in $q\bar{q}$-dominated $Z$ boson distribution.
A harder $q_{T}$ spectrum weakens the impact of the nonperturbative
Sudakov function \cite{Balazs:2000wv} and $q_{T}$ broadening. The
broadening in the rapidity-inclusive Higgs boson cross section is
further reduced because of the heavier Higgs boson mass and suppression
of the forward-rapidity regions, as displayed in~Fig.~\ref{fig:rap}(c).
The nonperturbative and $q_{T}$ broadening terms in $gg$-initiated
processes are not constrained yet by the data. In the following, we
evaluate the uncertainty in the Higgs boson $q_{T}$ distribution
by computing the resummed cross sections for several plausible choices
of $\rho(x)$ and ${\cal S}_{NP}(b,Q;b_{*})$. 

Fig.~\ref{fig:HLHC}(a) shows $q_{T}$ distributions for on-shell
Higgs bosons with a mass $M_{H}=120$ GeV and arbitrary rapidity.
We first compare the cross sections for $\rho(x)=0$ and $\rho(x)\neq0$
(thick lines), with the same functions ${\rho}(x)$ and ${\mathcal{S}}_{NP}(b,Q;\, b_{*})$
as in $Z$ boson production. The difference between the two curves
is minimal. The value of ${\cal A}^{(1)}$ reflects the dominant color
flow in the scattering. The bigger ${\cal A}^{(1)}$ in Higgs boson
production suggests that the $Q$-dependent part (and possibly other
terms) in ${\mathcal{S}}_{NP}$ may be multiplied by a larger color
factor than in the Drell-Yan process. To estimate this effect, we
multiply ${\mathcal{S}}_{NP}$ by $C_{A}/C_{F}=9/4$. The resulting
cross section (thin solid line) is close to the other cross sections.

The $\ln(1/x)$ terms in Higgs boson production may be enhanced as
well, due to the direct coupling of the Higgs bosons to gluon ladders.
However, the broadening would have to be quite large to affect $q_{T}$
of 20 GeV or more, i.e., near the location of the selection cuts on
$q_{T}$ of the photon pair. For example, increasing the function
$\rho(x)$ by a factor of two as compared to the $Z$ boson case would
lead to a distribution shown by the thin dashed line, which remains
close to the other cross sections at $q_{T}$ above 20 GeV. The broadening
is enhanced in the sample with $|y|>2$, which constitutes about 20\%
of the total cross section {[}Fig.~\ref{fig:HLHC}(b){]}. In this
figure, the differences between the curves are more substantial at
all $q_{T}$. The theoretical uncertainties seen in Fig.~\ref{fig:HLHC}(b)
will have to be reduced to reliably predict Higgs boson production
at forward rapidities.

The small-$x$ broadening will be also present in the main background
channels, $q\bar{q}\rightarrow\gamma\gamma$ and $gg\rightarrow\gamma\gamma$.
Application of the CSS formalism in these channels was discussed in
Refs.~\cite{Balazs:1997hv,Balazs:1999yf,Nadolsky:2002gj}. The dominant
background process, $q\bar{q}\rightarrow\gamma\gamma$, has the same
structure of the resummed cross section as $q\bar{q}\rightarrow Z^{0}$.
In analogy to $Z$ boson production, the broadening effect in $q\bar{q}\rightarrow\gamma\gamma$
is expected to be stronger than in $gg\rightarrow H^{0}$ or $gg\rightarrow\gamma\gamma$,
so that the selection cut on $q_{T}$ of the $\gamma\gamma$ pair
may have to be revised. Additional measurements will be needed to
constrain the small-$x$ behavior in the $gg$ channel, e.g., by examining
$\Upsilon$ production \cite{Kulesza:2003wi,Berger:2004cc} or $\gamma\gamma$
production away from the Higgs signal region. As the mass of the photon
pair decreases, the subprocess $gg\rightarrow\gamma\gamma$ becomes
increasingly important, contributing about 40\% of the total cross
section at $Q=80$ GeV at the LHC \cite{Bern:2002jx}. By comparing
$q_{T}$ distributions in $pp\rightarrow\gamma\gamma$ and $pp\rightarrow Z$
in the same region of $Q$, one may be able to separate the $q\bar{q}$
and $gg$ components of the resummed cross section and learn about
the $x$ dependence in the $gg$ channel.

\section{Conclusion}

In this paper, we analyze consequences of possible non-trivial rapidity
dependence in transverse momentum distributions of massive electroweak
bosons at the Tevatron and LHC. A specific mechanism considered here
assumes enhancement of radiative corrections in the impact-parameter-dependent
parton distributions at $x\lesssim10^{-2}$. This enhancement can
explain the semi-inclusive DIS data from HERA, and it leads to wider
transverse momentum distributions in the forward rapidity regions
at hadron colliders. A measurement of the rapidity dependence of $q_{T}$
distributions will test for the presence of such effects and verify
universality of $b$-dependent parton distributions, which was used
in this paper to relate the SIDIS and Drell-Yan cross sections. 

Due to the two-scale nature of the transverse momentum resummation,
large radiative corrections (such as non-resummed $\ln(1/x)$ terms)
may affect the $q_{T}$ distribution even when they leave no discernible
trace in the inclusive cross sections depending on one hard scale
$Q$. The Collins-Soper-Sterman resummed form factor depends on the
impact parameter $b,$ which specifies the factorization scale $\sim1/b$
in the $b$-dependent parton distributions. This scale changes from
zero to infinity when $b$ is integrated over the whole range in the
Fourier-Bessel transform to obtain the resummed $q_{T}$ distribution.
When $1/b$ is much smaller than $Q$, the series $\alphas^{m}(1/b)\ln^{n}(1/x)$
in the CSS formula diverge at a larger value of $x$ than the series
$\alphas^{m}(Q)\ln^{n}(1/x)$ in the inclusive cross sections. For
this reason, the transition to the small-$x$ dynamics may occur at
larger $x$ in $q_{T}$ distributions than in one-scale observables.

For a realistic choice of parameters, the rapidity-dependent $q_{T}$
broadening may be discovered via the analysis of forward-produced
$Z$ bosons at the Tevatron Run-2. The broadening could be also searched
for in the production of Drell-Yan pairs with a mass of a few GeV
in fixed-target and collider experiments. Due to the different masses
and shapes of the rapidity distributions for $W^{\pm},$ $Z^{0},$
and Higgs bosons, the magnitude of the broadening depends on the type
and charge of the boson produced, and selection cuts imposed on the
decay products. In the Tevatron Run-2, the resulting change may shift
the measured $W$ boson mass in the $p_{Te}$ method by $10-20$ MeV
in the central region ($\left|y_{e}\right|<1$) and more than $50$
MeV in the forward region ($\left|y_{e}\right|>1$). 

At the LHC, the most striking consequence is a harder $q_{T}$ distribution
for $W$ bosons displayed in Fig.~\ref{fig:WLHC}. The predicted
effect may easily exceed the other uncertainties in the resummed cross
section, resulting in important implications for the measurement of
the $W$ boson mass from both transverse mass and transverse momentum
distributions. The selection requirements imposed on the Higgs boson
candidates in the $\gamma\gamma$ decay channel may have to be reconsidered
to account for the non-uniform broadening in the signal and background
processes. More generally, the broadening may bring about a revision
of the earlier predictions that have neglected the correlation between
the transverse momentum and rapidity in Drell-Yan-like processes. 

\noindent \vskip 1cm

\section*{Acknowledgments}

We thank U.~Baur, R.~Kehoe, C.~R.~Schmidt, and R.~Stroynowski
for valuable discussions, and I.~Volobouev and G.~Steinbrueck for
the clarification of issues related to the CDF and \D0 detector acceptance.
F.~I.~O. and S.~B. acknowledge the hospitality of Fermilab and
BNL, where a portion of this work was performed. The work of S.~B.,
P.~M.~N., and F.~I.~O. at SMU was partially supported by the U.S.
Department of Energy under grant DE-FG03-95ER40908 and the Lightner-Sams
Foundation. The work of P.~M.~N. at Argonne National Laboratory
was supported in part by the US Department of Energy, High Energy
Physics Division, under Contract W-31-109-ENG-38. The work of C.-P.
Y. was partially supported by the National Science Foundation under
grant PHY-0244919.

\appendix

\section*{Appendix: Detector acceptances at the Tevatron and LHC}

In this appendix, we review the angular coverage of the hadronic detectors
and discuss feasibility of the observation of $q_{T}$ broadening
at the Tevatron and LHC.

In the Tevatron Run-2, adequate detector acceptance at large pseudorapidities
$\eta=\nolinebreak-\log(\tan(\theta/2))$ is crucial for the observation
of $q_{T}$ broadening in $Z$ boson production at $y\approx\left|\eta\right|\gtrsim2$.
$Z$ bosons can be identified by registering two high-$p_{T}$ electrons
in the electromagnetic calorimeter, with simultaneous positive identification
of at least one electron by the tracking system. In the \D0 detector,
the silicon tracking system extends out to $|\eta|\sim3$ \cite{Abachi:1994em,Hagopian2001,Ellison:2000sf},
which provides an excellent opportunity for collecting a clean sample
of forward-boosted $Z$ bosons in events with two registered forward
electrons.

At CDF \cite{Abe:1988me}, the unambiguous selection of the forward-boosted
$Z$ bosons is not possible due to the absence of tracking in the
forward region. The CDF detector can efficiently track charged electrons
in the main drift chamber at $|\eta|<1$ and in the silicon system
at $|\eta|<2$. Low efficiency tracking can be done up to $|\eta|\sim2.6$
by taking advantage of the longitudinally displaced collision vertex
in a fraction of events. Energy deposits from the electrons can be
registered in the plug calorimeter up to $|\eta|\sim3.6$. To increase
sensitivity in the forward region, CDF could concentrate on events
with one central ($\left|\eta\right|<2)$ and one forward ($\left|\eta\right|>2$)
electron. Despite contamination by central $Z$ bosons, this sample
may reveal the small-$x$ broadening once enough events are accumulated
in the next years of Run-2. 

The $q_{T}$ broadening in $Z$ boson production can be also detected
in muon decays. However, the coverage of the Tevatron muon systems
($|\eta|<1.5$ at CDF and $|\eta|<2$ at \D0) may be insufficient
in the kinematic region affected by the broadening. 

Excellent opportunities to probe the $x$ dependence of $q_{T}$ distributions
will become available at the LHC. Due to the larger center-of-mass
energy, the $q_{T}$ broadening may affect $W$ and $Z$ boson production
at all rapidities. In the ATLAS experiment, the tracking in the inner
detector extends up to $|\eta|\sim2.5$. The central and forward electromagnetic
calorimeters cover $|\eta|<4.9$, and the muon chambers cover $|\eta|<2.7\,$\cite{ATLAStdr:1999fq}.
In the CMS detector, the inner tracking system will register the high-$p_{T}$
charged particles in the range $|\eta|<2.5$. The electromagnetic
calorimeter extends up to $\left|\eta\right|=3$ \cite{cms:1997ki}.
Eventually, the TOTEM experiment will extend charged particle tracking
and triggering capabilities of the CMS experiment into the range $3<|\eta|<6.8\,$\cite{totem:Berardi:2004ku}.


\end{document}